\documentclass[%
 reprint,
superscriptaddress,
 amsmath,amssymb,
 aps,
 prx,
floatfix,
longbibliography,
]{revtex4-2}
\usepackage{graphicx}% Include figure files
\usepackage{array}[=2016-10-06]
\usepackage{dcolumn}% Align table columns on decimal point
\usepackage{bm}% bold math
\usepackage{color}
\usepackage{physics}
\usepackage{comment}
\usepackage{color}
\usepackage[dvipsnames]{xcolor}
\usepackage{ulem}
\usepackage{mathtools}
\usepackage{quantikz}
\usepackage{tikz}
\usepackage[colorlinks=true,linkcolor=blue,citecolor=blue,urlcolor=blue]{hyperref}
\usepackage{amsmath}
\usepackage{makecell}
\usepackage{ulem}

\begin{document}

\newcommand{\panel}[2]{%
\begin{tikzpicture}[baseline=(fig.base)]
\node[inner sep=0pt] (fig) {#2};
\node[
  anchor=north west,
  font=\bfseries,
  inner sep=1pt,
  xshift=-4mm,
  yshift=4mm
] at (fig.north west) {#1};
\end{tikzpicture}%
}
\tikzset{
  measureX/.style={
    label={[xshift=-0.3em,yshift=0.45em]below right:$X$}
  }
}
\preprint{APS/123-QED}
\title{Efficient Simulation of Hybrid Continuous- and Discrete-Variable Quantum Circuits via Gaussian Decompositions}% Force line breaks with \\

\author{Kazufumi Tanji}
\email[Contact author: ]{kizehemu@keio.jp}
\affiliation{%
 Department of Electronics and Electrical Engineering, Faculty of Science and Engineering, Keio University, Yokohama, Kanagawa 223-8522, Japan
}%
\author{Dot Belin Pio}
\affiliation{%
 Center for Macroscopic Quantum States (bigQ), Department of Physics, Technical University of Denmark, Building 307, Fysikvej, 2800 Kgs. Lyngby, Denmark
}%

\author{Shohei Kiryu}
\affiliation{%
 Department of Electronics and Electrical Engineering, Faculty of Science and Engineering, Keio University, Yokohama, Kanagawa 223-8522, Japan
}%

\author{Ulrik Lund Andersen}
\affiliation{%
 Center for Macroscopic Quantum States (bigQ), Department of Physics, Technical University of Denmark, Building 307, Fysikvej, 2800 Kgs. Lyngby, Denmark
}%

\author{Masahiro Takeoka}
\email[Contact author: ]{takeoka@elec.keio.ac.jp}
\affiliation{%
 Department of Electronics and Electrical Engineering, Faculty of Science and Engineering, Keio University, Yokohama, Kanagawa 223-8522, Japan
}%
\affiliation{Advanced ICT Research Institute, National Institute of Information and
Communications Technology (NICT), Koganei, Tokyo 184-8795, Japan}
\date{\today}

\begin{abstract}
Hybrid quantum systems combining discrete-variable (DV) and continuous-variable (CV) subsystems arise across a broad range of physical platforms and enable quantum information processing beyond purely qubit-based designs. Their classical simulation, however, is challenging since computational costs of both CV and DV systems exponentially increase. In this paper, we introduce a simulation framework for CV--DV hybrid systems based on linear combination of Gaussian (LCoG) decomposition. We decompose the hybrid density matrix into blocks in a DV basis and represent each block as an LCoG expansion. Within the representation, Gaussian operations on the bosonic subsystem independently act on each Gaussian function. We derive analytic formulas describing how complex Gaussian functions transform under controlled Gaussian operations. We further extend the framework to controlled Gaussian channels with state-dependent Gaussian noise. The method avoids a Fock-space truncation for the CV subsystem, and the cost of updating each Gaussian term scales polynomially with the number of modes. The framework provides a general computational tool for analyzing hybrid quantum algorithms, quantum simulation protocols, and non-Gaussian state generation schemes. As an example, we simulate Gottesman--Kitaev--Preskill (GKP) state generation based on a cavity-QED system.
\end{abstract}

\maketitle

\section{Introduction}
\begin{figure*}[t]
    \centering
    \includegraphics[width=0.9\linewidth]{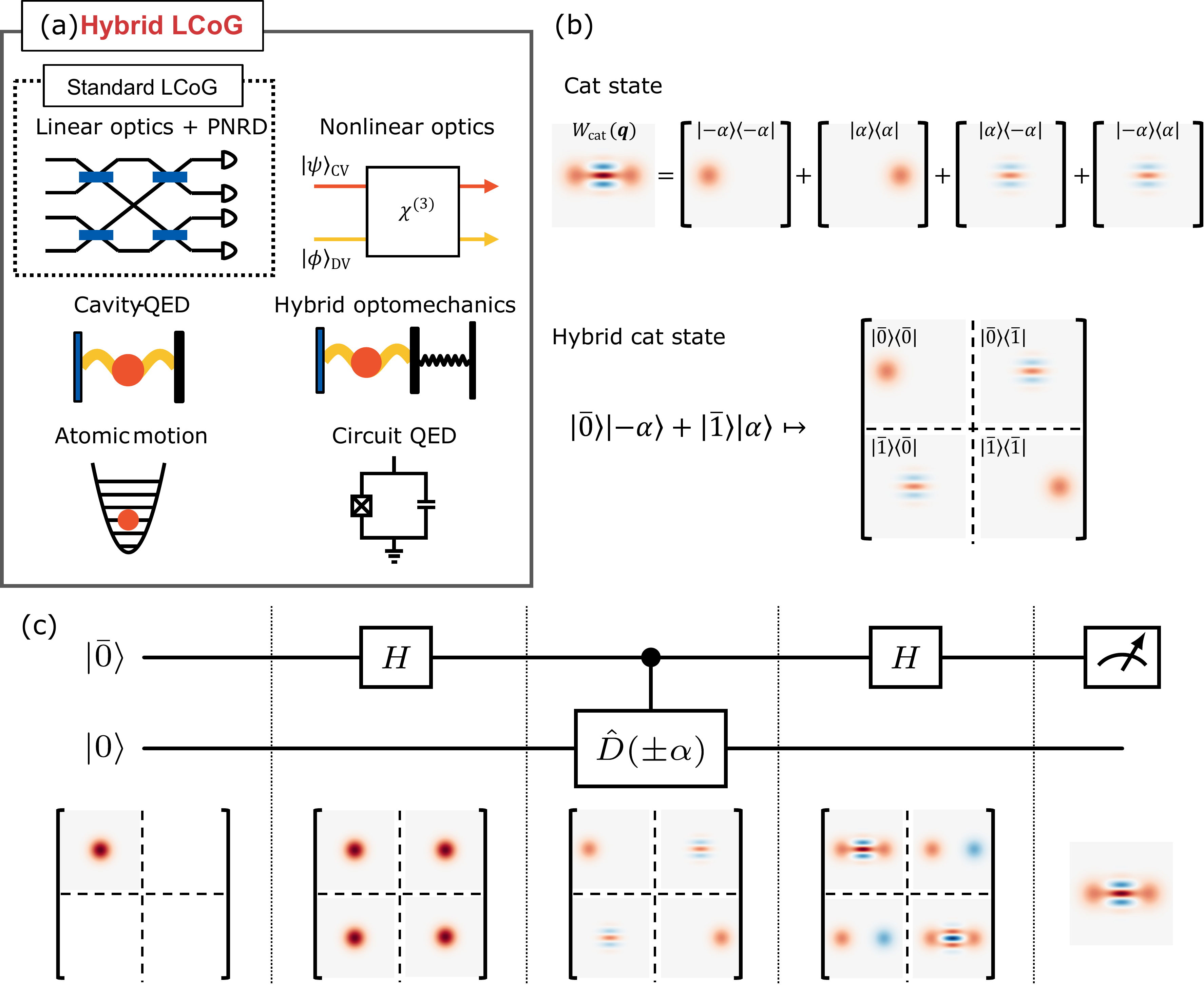}
    \caption{Summary of hybrid LCoG formalism. (a) Target physical platforms: linear and nonlinear photonic systems, cavity-QED, atomic motion, hybrid optomechanical system, and circuit QED system. (b) Examples of LCoG decomposition of a cat state (upper panel) and hybrid LCoG decomposition of a hybrid cat state (lower panel). The cat state is decomposed into two diagonal terms ($\ket{\pm\alpha}\bra{\pm\alpha}$) and two off-diagonal terms ($\ket{\pm\alpha}\bra{\mp\alpha}$). In phase space, the diagonal terms are represented by a real Gaussian function, while the off-diagonal terms are represented by complex Gaussian functions. Only real parts are shown here. In the hybrid cat state, the Gaussian functions are assigned to the corresponding qubit blocks. (c) Example circuit of cat state generation (upper panel) and evolution of the state in hybrid LCoG representation (lower panel).}
    \label{fig:summary}
\end{figure*}

Discrete-variable (DV) and continuous-variable (CV) quantum systems coexist and interact in a wide range of physical platforms~\cite{andersenHybridDiscreteContinuousvariable2015b,kemperHybridContinuousdiscretevariableQuantum2025,liuHybridOscillatorQubitQuantum2026b}, including  trapped ions~\cite{bruzewiczTrappedionQuantumComputing2019}, photonic systems~\cite{andersenHybridDiscreteContinuousvariable2015b}, circuit and cavity-QED systems~\cite{blaisCircuitQuantumElectrodynamics2021}, and optomechanical systems~\cite{aspelmeyerCavityOptomechanics2014} as shown in Fig.~\ref{fig:summary}(a). A central operation is a controlled Gaussian operation, where the Gaussian transformation applied to a CV system depends on the state of the DV system. Such operations have been proposed or demonstrated in trapped ions~\cite{bruzewiczTrappedionQuantumComputing2019}, superconducting circuits~\cite{blaisCircuitQuantumElectrodynamics2021}, nonlinear optics~\cite{munroWeakNonlinearitiesNew2005,yamaguchiQuantumErrorCorrection2006,myersStabilizerQuantumError2007,vanloockHybridQuantumComputation2008,nemotoNearlyDeterministicLinear2004}, and cavity-QED systems~\cite{kikuraSingleshotConditionalDisplacement2025,hackerDeterministicCreationEntangled2019}. They underpin the preparation and control of bosonic codes, including cat codes~\cite{leghtasHardwareEfficientAutonomousQuantum2013,hackerDeterministicCreationEntangled2019,notarnicolaDeterministicFeedforwardbasedGeneration2026,kikuraEngineeringPropagatingCat2025,hoshiEntanglingSchrodingersCat2025,vlastakisCharacterizingEntanglementArtificial2015}, binomial codes~\cite{michaelNewClassQuantum2016a,niBeatingBreakevenPoint2023,tejaCavityQEDToolsMBQC2026,tejaQuantumNonGaussianHigh2025,matsosRobustDeterministicPreparation2024a}, and Gottesman--Kitaev--Preskill (GKP) codes~\cite{gottesmanEncodingQubitOscillator2001a,hastrupProtocolGeneratingOptical2022a,hastrupMeasurementfreePreparationGrid2021a,fluhmannEncodingQubitTrappedion2019a,deneeveErrorCorrectionLogical2022a,matsosRobustDeterministicPreparation2024a,matsosUniversalQuantumGate2025,brockQuantumErrorCorrection2025,grimsmoQuantumErrorCorrection2021c,sivakRealtimeQuantumError2023,millicanEngineeringContinuousVariableEntanglement2025}. 
Controlled displacements also realize geometric-phase gates such as the M{\o}lmer-S{\o}rensen gate~\cite{molmerMultiparticleEntanglementHot1999,sorensenQuantumComputationIons1999}, which is a standard multi-qubit gate on trapped ion quantum computers~\cite{ransfordHelios98qubitTrappedion2025,hughesTrappedionTwoqubitGates2025,katz$N$BodyInteractionsTrapped2022,katzDemonstrationThreeFourbody2023a}. Similar gates have been developed in superconducting circuits~\cite{royerFastHighfidelityEntangling2017} and cavity-QED systems~\cite{takahashiMolmerSorensenEntangling2017}, as well. More broadly, these hybrid operations enable high-precision quantum sensing with less overhead~\cite{valahuQuantumenhancedMultiparameterSensing2025,bondOptimalDisplacementSensing2026a,sinanan-singhSingleshotQuantumSignal2024,panRealizationVersatileEffective2025a,zhengQuantumenhancedSensingBosonic2025a} and long-distance quantum communications~\cite{vanloockQuantumRepeatersUsing2008,bergmannHybridQuantumRepeater2019,liMemorylessQuantumRepeaters2023,guccioneConnectingHeterogeneousQuantum2020,liPerformanceRotationSymmetricBosonic2024}.

Beyond these elementary operations, CV–DV hybrid systems provide distinct architectures for quantum computation and simulation. Compared with purely DV encodings, hybrid architectures are advantageous for simulating quantum chemistry~\cite{huhBosonSamplingMolecular2015a,huSimulationMolecularSpectroscopy2018,sparrowSimulatingVibrationalQuantum2018a,wangObservationWavePacketBranching2023,katzProgrammableQuantumSimulations2023,olaya-agudeloSimulatingOpensystemMolecular2025,wangEfficientMultiphotonSampling2020,bellCodesigningSpectralTransformation2025} and lattice gauge theories~\cite{craneHybridOscillatorQubitQuantum2024,sanerRealTimeObservationAharonovBohm2025} since their continuous- and discrete-variables can be encoded directly into bosonic modes and qubits, respectively. The same architecture has also motivated hybrid quantum algorithms~\cite{hastrupUniversalUnitaryTransfer2022,liuMixedAnalogDigitalQuantum2025,brennerFactoringIntegerThree2025,duttaSolvingConstrainedOptimization2026,dasQuantumDifferentialEquation2026} and fault-tolerant quantum computer architectures~\cite{royerEncodingQubitsMultimode2022,nohLowOverheadFaultTolerantQuantum2022,leeFaultTolerantQuantumComputation2024}.

Classical simulation is essential for analyzing and benchmarking these versatile applications, but remains computationally demanding. Since a CV system has an infinite-dimensional Hilbert space, it is commonly represented in a Fock basis truncated at a finite photon-number cutoff~\cite{stavengerC2QABosonicQiskit2022,mohapatraHyQBenchBenchmarkSuite2026a,furchesHybridlaneSoftwareDevelopment2026c}. The required cutoff grows rapidly with the energy introduced by displacement and squeezing, scaling quadratically with the displacement amplitude and exponentially with the squeezing parameter. For a multimode system, the dimension of the resulting truncated Hilbert space increases exponentially with the number of modes. Incorporating a DV system introduces an additional exponential dependence on the number of qubits. Moreover, an insufficient cutoff causes truncation errors that are difficult to estimate from the circuit parameters alone. This issue is particularly important for variational quantum algorithms~\cite{duttaSolvingConstrainedOptimization2026}, in which displacement and squeezing parameters change during optimization. Consequently, existing general-purpose simulators are restricted to relatively small CV–DV systems~\cite{stavengerC2QABosonicQiskit2022,mohapatraHyQBenchBenchmarkSuite2026a,furchesHybridlaneSoftwareDevelopment2026c}.

These limitations can be alleviated by restricting the classes of states and operations. In DV systems, stabilizer states evolving under Clifford operations can be simulated in polynomial time~\cite{gottesmanHeisenbergRepresentationQuantum1998}. Likewise, Gaussian states evolving under Gaussian operations can be simulated efficiently in CV systems~\cite{weedbrookGaussianQuantumInformation2012}. A Gaussian Wigner function is completely specified by its mean vector and covariance matrix. However, the stabilizer and Gaussian formalisms are restricted to Clifford and Gaussian operations, respectively~\cite{liuHybridOscillatorQubitQuantum2026b}. Controlled Gaussian operations are generally neither Clifford nor Gaussian and therefore fall outside both formalisms.

Recently, the Gaussian formalism has been extended to include non-Gaussian states through the linear combination of Gaussians (LCoG) formalisms~\cite{bourassaFastSimulationBosonic2021,solodovnikovaFastSimulationsContinuousvariable2025a} as shown in the upper panel of Fig.~\ref{fig:summary}(b). In this formalism, a non-Gaussian state is represented as a linear combination of Gaussian functions in phase space. Gaussian operations and channels preserve this form since they transform each Gaussian function into another Gaussian function. The coherent-state decomposition further enables approximate LCoG representations of Fock states and photon-number-resolving detection (PNRD)~\cite{marshallSimulationQuantumOptics2023}, allowing efficient simulations of non-Gaussian state-preparation protocols for optical bosonic codes~\cite{takedaDeterministicQuantumTeleportation2013,morinRemoteCreationHybrid2014,takedaEntanglementSwappingDiscrete2015,kiryuLinearOpticalQuantum2025,kiryuLinearopticalGenerationHybrid2026,konnoLogicalStatesFaulttolerant2024,larsenIntegratedPhotonicSource2025}. Although PNRD increases the number of Gaussian terms, the resulting simulation remains substantially more efficient than a direct Fock-basis expansion. Despite this progress, extending the LCoG formalism to CV–DV hybrid systems is nontrivial since DV and CV states are described in fundamentally different ways. Therefore, it remains unclear how hybrid states should be represented and how operations should be described within the formalism.

In this paper, we develop a hybrid LCoG formalism for CV–DV systems. Our main contributions are as follows.
First, we introduce a representation of hybrid states in which the hybrid density operator is written as a block matrix in the DV computational basis and each block is expanded in LCoG form as illustrated in the lower panel of Fig.~\ref{fig:summary}(b).
Within this representation, we can treat arbitrary qubit gates, channels, and measurements on the DV subsystem, together with Gaussian operations and channels, Gaussian measurements, and PNRD on the CV subsystem. Qubit operations mix or select different DV blocks, whereas CV operations transform the Gaussian terms within each block, as shown in Fig.~\ref{fig:summary}(c). We also give the corresponding update rules,
resulting numbers of Gaussian terms, and computational costs. Second, we derive closed-form transformation rules for controlled Gaussian unitaries, which constitute the central theoretical challenge of the formalism. For an off-diagonal block, two different Gaussian unitaries act on the CV operator from the left and right, and the usual transformation rule for Gaussian states therefore does not apply. We show that each complex Gaussian function nevertheless remains Gaussian, and derive its transformed complex weight, mean vector, and covariance matrix. This result allows controlled Gaussian operations to be evaluated entirely within the hybrid LCoG representation. Third, using a Stinespring dilation, we extend these rules to controlled Gaussian channels whose loss and noise depend on the DV state.
The resulting simulation requires no finite-dimensional truncation of the CV Hilbert space. For both Gaussian and controlled Gaussian operations, the computational cost scales linearly with the number of Gaussian terms and polynomially with the number of bosonic modes.

These results extend the LCoG formalism from purely CV circuits to a broad range of CV–DV applications, including non-Gaussian state generation, hybrid quantum algorithms, quantum simulation, and the benchmarking of hybrid devices. Beyond numerical simulation, the transformation formulas provide an analytic description of how controlled Gaussian operations generate and transform non-Gaussian states. To demonstrate its performance, we simulate GKP state generation in a cavity-QED system~\cite{hastrupProtocolGeneratingOptical2022a}.

\textit{Note added.} After the completion of this work, version 2 of Ref.~\cite{braccini2026superpositionsquantumgaussianprocesses} was posted on September 23, 2026. Compared with the original version, this updated version additionally describes analytical maps for controlled displacements and rotations, as well as qubit operations that generate sums of Gaussians within each block. These results partially overlap with Secs.~III and IV D 1--2 of the present work.

\section{Preliminaries}
In this section, we introduce the phase-space notation and the linear-combination-of-Gaussians (LCoG) representation. 
In Sec.~{\ref{sec:2.A}}, we define the quadrature operators and the Wigner function. In Sec.~\ref{sec:2.B}, we describe Gaussian operations and measurements in phase space. In Sec.~\ref{sec:2.C}, we review the LCoG formalism introduced in Refs.~\cite{bourassaFastSimulationBosonic2021,solodovnikovaFastSimulationsContinuousvariable2025a}.

\subsection{\label{sec:2.A}Wigner functions}
We consider an $M$-mode system with $2M$ quadrature operators represented by the vector $\hat{\bm q}=[\hat x_1, \hat p_1,\ldots,\hat x_M,\hat p_M]^T$. These quadrature operators satisfy the canonical commutation relations $[\hat q_i,\hat q_j]=i\Omega_{ij}$, where $\Omega$ is the $2M\times 2M$ symplectic form:
\begin{align}
    \Omega=\bigoplus_{i=1}^M\left(\begin{matrix}
        0&1\\
        -1&0
    \end{matrix}\right).
\end{align}

An $M$-mode quantum state $\hat \rho$ is represented in phase space by a Wigner function, which is defined by the Weyl transform:
\begin{align}
    \label{eq:Weyl_transform}
        W_{\hat \rho}(\bm q)=\frac{1}{(2\pi^2)^{M}}\int d^{2M}\bm \zeta\ \Tr[\hat\Theta_{\bm q,\bm \zeta}\hat \rho]e^{i\bm q^T \bm \Omega\bm \zeta},
\end{align}
where $\hat\Theta_{\bm q,\bm \zeta}$ is an outer product of two different coherent states:
\begin{align}
    \label{eq:definition_of_Pi_q_zeta}
    \hat \Theta_{\bm q,\bm \zeta}=\hat D\left(\frac{\bm q-\bm\zeta}{\sqrt 2}\right)\ket{0}\bra 0\hat D^\dagger\left(\frac{\bm q+\bm \zeta}{\sqrt 2}\right)
\end{align}
with the Weyl operator
\begin{equation}
    \label{eq:Weyl_operator}
    \hat D(\bm \alpha)=\exp(i\sqrt{2}\hat{\bm q}^T\bm \Omega\bm \alpha).
\end{equation}

A quantum state $\hat \rho$ is Gaussian if and only if its Wigner function is a multivariate Gaussian function:
\begin{equation}
    \begin{aligned}
        W_{\hat\rho}(\bm q)
        &=\frac{1}{\sqrt{\det 2\pi\bm{\sigma}}}\exp\left[-\frac{1}{2}(\bm q-\bm \mu)^T\bm{\sigma}^{-1}(\bm q-\bm \mu)\right]\\
        &\eqqcolon G_{\bm \mu,\bm{\sigma}}(\bm q),
    \end{aligned}
\end{equation}
where $\bm \mu$ and $\bm{\sigma}$ denote the mean vector and the covariance matrix, respectively, defined as
\begin{equation}
    \begin{aligned}
        \mu_i&=\langle\hat q_i\rangle,\\
        \bm{\sigma}_{ij}&=\frac{1}{2}\langle\{\hat q_i-\langle \hat q_i\rangle,\hat q_j-\langle \hat q_j\rangle\}\rangle,
    \end{aligned}
\end{equation}
Here, $\{\cdot,\cdot\}$ denotes the anti-commutator. The covariance matrix must satisfy the uncertainty relation, $\bm{\sigma}\geq -i\frac{1}{2}\bm \Omega$.

\subsection{\label{sec:2.B}Gaussian operations}
A unitary operator $\hat U_G$ is Gaussian unitary if it maps every Gaussian state to another Gaussian state. A Gaussian unitary acting on quadrature operators is described by a symplectic matrix $S$ and a displacement vector $\bm \alpha$:
\begin{align}
    \hat U_G^\dagger \hat{\bm q}\hat U_G=S\bm {\hat q}+\bm \alpha,
\end{align}
Consequently, the mean vectors and covariance matrices transform as
\begin{equation}
    \label{eq:gaussian_unitary_update}
    \begin{alignedat}{2}
        \bm \mu_\text{out}&= S\bm \mu_\text{in}+\bm \alpha,
        \qquad&
        \bm{\sigma}_\text{out}&= S\bm{\sigma}_\text{in} S^T.
    \end{alignedat}
\end{equation}

A Gaussian channel is a completely positive and trace-preserving (CPTP) map that maps a Gaussian state to another Gaussian state. Its action on the mean vector and the covariance matrix is described by real matrices $X$ and $Y$, together with a real displacement vector $\bm \alpha$, as~\cite{weedbrookGaussianQuantumInformation2012}
\begin{equation}
    \label{eq:gaussian_channel_update}
    \begin{alignedat}{2}
        \bm \mu_\text{out}&=\bm X\bm \mu_\text{in}+\bm \alpha,
        \qquad&
        \bm{\sigma}_\text{out}&=\bm X\bm{\sigma}_\text{in} \bm X^T+\bm Y.
    \end{alignedat}
\end{equation}
Complete positivity requires $\bm Y+\frac{i}{2}\bm \Omega\geq\frac{i}{2}\bm X\bm \Omega \bm X^T$. Gaussian unitaries correspond to the case $X=S$ and $Y=0$.

Consider a measurement described by the positive operator-valued measure (POVM) element $\hat{\mathcal M}(\mathbf{x})$ with the measurement outcome $\mathbf{x}$. The probability density for obtaining $\mathbf x$ is
\begin{align}
    \label{eq:trace}
    p(\mathbf x)=\Tr[\hat \rho \hat{\mathcal M}(\mathbf x)].
\end{align}
A measurement is Gaussian when the Weyl transform of each POVM element is Gaussian. Using the phase-space trace rule, we obtain
\begin{align}
    \Tr[\hat \rho \hat{\mathcal M}(\mathbf x)]=(2\pi)^M\int d^{2M}\bm q\ W_{\hat \rho}(\bm q)W_{\hat{\mathcal M}(\mathbf x)}(\bm q)
\end{align}
For a Gaussian state with Wigner function $G_{\bm \mu,\bm{\sigma}}(\bm q)$, the outcome probability density is
\begin{equation}
    \label{eq:trace_Gaussian}
    \begin{aligned}
        p(\mathbf x)&=\int d^{2M}\bm q\ G_{\bm \mu,\bm{\sigma}}(\bm q)G_{\mathbf x,\bm{\sigma}_{\hat{\mathcal M}}}(\bm q)\\
        &=G_{\bm \mu, \bm{\sigma}+\bm{\sigma}_{\hat{\mathcal M}}}(\mathbf x).
    \end{aligned}
\end{equation}

We next consider a composite Gaussian state on $\mathcal H=\mathcal H_A\otimes \mathcal{H}_B$, where the subsystem $B$ is measured.
Its mean vector and the covariance matrix are partitioned as
\begin{equation}
    \begin{alignedat}{2}
        \bm \mu&=\begin{bmatrix}
                \bm \mu_A\\
                \bm \mu_B
            \end{bmatrix},
        \qquad&
        \bm{\sigma}&=\begin{bmatrix}
            \bm{\sigma}_A & \bm{\sigma}_{AB} \\
            \bm{\sigma}_{AB}^T & \bm{\sigma}_B
        \end{bmatrix}.
    \end{alignedat}
\end{equation}
Suppose that a Gaussian measurement with outcome $\mathbf x$ and covariance matrix $\bm{\sigma}_{\hat{\mathcal{M}}}$ is performed on subsystem $B$. The Wigner function of the unnormalized conditional state on subsystem A is given by integrating over the vector $\bm q_B$:
\begin{equation}
        \Tr_B[\hat \rho_{AB}\hat{\mathcal{M}}_B(\mathbf x)]\mapsto G_{\bm \mu_B, \bm{\sigma}_B+\bm{\sigma}_{\hat{\mathcal M}}}(\mathbf x)G_{\bm \mu_A^\text{out}(\mathbf x),\bm{\sigma}_A^\text{out}}(\bm q_A).
\end{equation}
The first Gaussian factor gives the probability density of the measurement outcome, while the second is the normalized Wigner function of the conditional state. Its covariance matrix and mean vector are given by
\begin{equation}
    \label{eq:partial_gaussian_measurment_update}
        \begin{aligned}
            \bm{\sigma}_A^\text{out}&=\bm{\sigma}_A-\bm{\sigma}_{AB}(\bm{\sigma}_B+\bm{\sigma}_{\hat{\mathcal M}})^{-1}\bm{\sigma}_{AB}^T,\\
            \bm \mu_A^\text{out}(\mathbf x)&=\bm \mu_{A}+\bm{\sigma}_{AB}(\bm{\sigma}_B+\bm{\sigma}_{\hat{\mathcal M}})^{-1}(\mathbf x-\bm \mu_B).
        \end{aligned}
\end{equation}

\subsection{\label{sec:2.C}LCoG formalisms}
In the previous section, we restricted our attention to Gaussian states.
Hereafter, we consider a broader class of quantum states that can be decomposed into linear combination of Gaussians:
\begin{equation}
    \label{eq:decomposition_of_density_matrix}
    \begin{aligned}
        \hat\rho&=\sum_{i,j} p_{ij}\ket{G_i}\bra{G_j},
    \end{aligned}
\end{equation}
where $\ket{G_i}$ represents a pure Gaussian state.
Applying the Weyl transform to an individual outer product gives
\begin{equation}
    \begin{aligned}
        &W_{\ket{G_i}\bra{G_j}}(\bm q)\\
        =&\frac{1}{(2\pi^2)^{M}}\int d^{2M}\bm \zeta\ \Tr[\hat \Theta_{\bm q,\bm \zeta}\ket{G_i}\bra{G_j}]e^{i\bm q^T\bm \Omega\bm \zeta},
\end{aligned}
\end{equation}
The resulting function is a complex Gaussian:
\begin{align}
    W_{\ket{G_i}\bra{G_j}}(\bm q)=e^{d}G_{\bm \mu,\bm\sigma}(\bm q),
\end{align}
where $d$ denotes the log-weight, $\bm \mu$ denotes the mean vector, and $\bm \sigma$ denotes the covariance matrix.

For example, we consider the outer product of two different coherent states $\ket{\bm \alpha_1}\bra{\bm \alpha_2}$ with the amplitudes $\bm \alpha_1$ and $\bm \alpha_2$. Its Gaussian parameters are derived in Ref.~\cite{bourassaFastSimulationBosonic2021} as
\begin{equation}
    \label{eq:params_outer_product_of_coherent_state}
    \begin{aligned}
        \bm \mu&=\Pi_0\bm \alpha_1+\Pi_1 \alpha_2\\
        \bm \sigma &=\frac{1}{2}\mathbb I\\
        d&=\bm \alpha_1^T\Pi_1\bm \alpha_2-\frac{\|\bm \alpha_1\|^2+\|\bm\alpha_2\|^2}{4},
    \end{aligned}
\end{equation}
where 
\begin{align}
    \label{eq:projector_Omega}
    \Pi_k=\frac{\mathbb I+(-1)^k i\bm \Omega}{2}\quad (k\in\{0,1\})
\end{align}
are projection operators onto the $\pm 1$ eigenspaces of $i\bm \Omega$.

Applying Weyl transform for all outer products in Eq.~(\ref{eq:decomposition_of_density_matrix}), the Wigner function of $\rho$ is decomposed into the LCoG form
\begin{align}
    \hat \rho\rightarrow W_{\hat \rho}(\bm q)=\sum_{n} e^{d_n} G_{\bm \mu_n, \bm{\sigma}_n}(\bm q),
\end{align}
where $\sum_n e^{d_n}=1$ follows the normalization condition of the density operator. Gaussian channels act independently on each term according to the transformations in Eq.~(\ref{eq:gaussian_channel_update}). In particular, under a Gaussian unitary $\hat U_G$, we obtain
\begin{align}
    W_{\hat U_G\hat \rho\hat U_G^\dagger}(\bm q)=\sum_{n\in \mathbb N} e^{d_n} G_{S\bm \mu_n+\bm \alpha, S\bm \sigma_nS^T}(\bm q),
\end{align}
where the symplectic matrix $S$ and the mean vector $\bm \alpha$ correspond to the given Gaussian unitary $\hat U_G$.

States that do not admit an exact LCoG representation can instead be approximated by Gaussian decomposition. 
Marshall and Anand introduced a coherent-state decomposition in which a Fock state $\ket{n}$ is approximated by an $(n+1)$-legged cat state with complex weights~\cite{marshallSimulationQuantumOptics2023}:
\begin{align}
    \ket n\approx \frac{1}{\sqrt{\mathcal N_n}}\sum_{k=0}^{n}e^{-ikn\theta_n}\ket{\bm\alpha_k},
\end{align}
where $\sqrt{\mathcal N_n}$ is the normalization factor, and the displacement vector is given by
\begin{equation}
    \begin{alignedat}{2}
        \bm\alpha_k&=\frac{\epsilon}{\sqrt 2} \begin{bmatrix}
            \cos k\theta_n\\
            \sin k\theta_n
        \end{bmatrix},
        &\qquad
        \theta_n&=\frac{2\pi}{n+1}
    \end{alignedat}
\end{equation}
with free parameter $\epsilon$, controlling accuracy.

\section{Hybrid LCoG States}
In this section, we extend the LCoG representation to hybrid systems composed of qubits and bosonic modes. In Sec.~\ref{sec:3.A}, we introduce the representation for a single qubit and a single bosonic mode, and in Sec.~\ref{sec:3.B}, we generalize it to multiple qubits and modes.

\subsection{\label{sec:3.A} Single-qubit and single-mode systems}
We consider a hybrid system composed of a single qubit and a single bosonic mode, with Hilbert space $\mathcal H_\text{q}\otimes \mathcal H_\text{b}$.
A general qubit-oscillator state is given by
\begin{align}
    \label{eq:single-qubit-oscillator-state}
    \hat \rho_\text{q,b}=\left(\begin{matrix}
        \hat \rho_{00} & \hat \rho_{01}\\
        \hat \rho_{10} & \hat \rho_{11}
    \end{matrix}\right),
\end{align}
where $\hat\rho_{jk}$ is an operator on $\mathcal{H}_\text{b}$.

When the block matrix $\rho_{jk}$ can be decomposed into the LCoG form, its Weyl transform takes the form
\begin{equation}
    W_{\rho_{jk}}(\bm q)=\sum_{l=0}^{L_{jk}-1}e^{d_{jkl}}G_{\bm\mu_{jkl},\bm{\sigma}_{jkl}}(\bm q),
\end{equation}
where $L_{jk}$ is the number of Gaussian terms in the $(j,k)$ block.
From the trace preservation of the density matrix, the log-weights must satisfy $\sum_{l}\exp(d_{00l})+\exp(d_{11l})=1$. Therefore, the full state can be represented by the list $\{d_{jkl}, \bm \mu_{jkl}, \bm{\sigma}_{jkl}\}$.

\subsection{\label{sec:3.B} Multi-qubit and multi-mode systems}
We next consider a system composed of $N$ qubits and $M$ modes on $\mathcal H_A\otimes \mathcal H_B$, where 
\begin{equation}
    \begin{alignedat}{2}
        \mathcal H_A&=\mathcal{H}_\text{q}^{\otimes N},
        &\qquad
        \mathcal H_B&= \mathcal{H}_\text{b}^{\otimes M}
    \end{alignedat}
\end{equation}
For notational simplicity, we introduce binary strings to represent the $N$-qubit computational basis, i.e., $\ket{\bar{\bm j}}=\ket{\bar j_0,\ldots,\bar j_{N-1}}$.

Then, an arbitrary state on $\mathcal H_A\otimes \mathcal H_B$ can be written as
\begin{equation}
    \begin{aligned}
        \hat\rho_{AB}=&\sum_{\bm j,\bm k\in\{0,1\}^N}\ket{\bar{\bm j}}\bra{\bar{\bm k}}\otimes\hat\rho_{\bm j\bm k},
    \end{aligned}
\end{equation}
where $\hat \rho_B^{\bm j\bm k}$ denotes the operator on $\mathcal H_B$.

When every operator $\hat \rho_{B}^{\bm j\bm k}$ can be decomposed into the LCoG form, its Weyl transform is
\begin{equation}
    W_{\hat \rho_{\bm j \bm k}}(\bm q)=\sum_{l=0}^{L_{\bm j\bm k}-1}e^{d_{\bm j\bm kl}}G_{\bm\mu_{\bm j\bm kl},\bm{\sigma}_{\bm j\bm kl}}(\bm q).
\end{equation}
From the trace preservation of the density matrix, the log-weights must satisfy $\sum_{\bm j}\sum_{l}\exp(d_{\bm j\bm jl})=1$. Therefore, the full state can be represented by the list of Gaussian parameters and their log-weights, $\{d_{\bm j\bm kl}, \bm \mu_{\bm j\bm kl}, \bm{\sigma}_{\bm j\bm kl}\}$.

The memory cost of this representation is characterized by the total number of Gaussians $N_L=\sum_{\bm j,\bm k}L_{\bm j\bm k}$. Since each covariance matrix contains $2M\times 2M$ entries, the memory cost scales as $\mathcal{O}(N_L M^2)$. Specifically, if every block contains the same number $L$ of Gaussian functions, the memory cost scales as $\mathcal{O}(4^NLM^2)$. Here, the factor of $4^N$ comes from the number of blocks corresponding to $N$-qubit density matrix.

\section{\label{sec:4}Operations on Hybrid LCoG States}
In this section, we describe how quantum operations act on hybrid LCoG states, including unitary operators, quantum channels, and measurements. 
In Sec.~\ref{sec:4.A}, we summarize operations considered in this work and their computational costs.
In Sec.~\ref{sec:4.B}, we recall the Gaussian channels and measurements introduced in Sec.~\ref{sec:2.B} and apply them to the hybrid LCoG state. 
In Sec.~\ref{sec:4.C}, we introduce qubit operations on the hybrid LCoG states and estimate their computational cost. 
In Sec.~\ref{sec:4.D}, we derive the transformation rules for controlled Gaussian operations and controlled Gaussian channels.

\subsection{\label{sec:4.A} Overview and scaling}
As introduced in Sec.~\ref{sec:3.B}, the hybrid LCoG state is represented by the log-weights, mean vectors, and covariance matrices $\{d_{\bm j\bm kl},\bm \mu_{\bm j\bm kl},\bm{\sigma}_{\bm j\bm kl}\}$. We use the total number of Gaussians denoted by $N_L$ to characterize the computational cost. 

In the rest of Sec.~\ref{sec:4}, we consider the following three classes of operations: bosonic operations, qubit operations, and controlled Gaussian operations. Bosonic operations involve Gaussian channels, Gaussian (partial) measurements, and PNRD. For qubit operations, we classify them into monomial gates, general $n$-qubit gates and channels, and measurements. The monomial gate is defined by a unitary matrix with exactly one nonzero entry in each row and each column. Controlled Gaussian operations comprise controlled Gaussian unitaries and controlled Gaussian channels.

\begin{table*}[htbp]
    \caption{Computational cost for operations on hybrid LCoG states. $N_L$: the number of Gaussian terms in the input state. $N_\text{PNRD}$: the number of Gaussian functions used to approximate a Fock state. $N_K$: the number of Kraus operators. $M$: the number of modes in the system. $M_E$: the number of modes in environment.}
    \label{tab:operations_hybrid_LCoG}
    \begin{ruledtabular}
        \begin{tabular}{c|cccc}
            Operations & Log-weight $d$ & Gaussian parameters $\mu, \ \bm{\sigma}$ & \# of Gaussians & Computational costs\\\hline
            Gaussian channels & unchanged &updated & $N_L$ & $\mathcal{O}(N_L M^3)$\\
            Gaussian measurements & updated &updated & $N_L$ & $\mathcal{O}(N_L M^3)$\\
            PNRD & updated & updated & $\leq N_\text{PNRD}N_L$ & $\mathcal{O}(N_\text{PNRD}N_L M^3)$\\
            $n$-qubit monomial gates & updated & unchanged & $N_L$ & $\mathcal O(N_L)$\\
            $n$-qubit channels & updated & unchanged & $\leq 4^n N_KN_L$ & $\mathcal O(4^n N_K N_L)$\\
            Qubit measurements & updated & unchanged & $\leq N_L$ & $\mathcal O(N_L)$\\
            State-dependent Gaussian channels & updated & updated & $N_L$ & $\mathcal{O}(N_L (M+M_E)^3)$\\
        \end{tabular}
    \end{ruledtabular}
\end{table*}

Table~\ref{tab:operations_hybrid_LCoG} summarizes the elementary operations on hybrid LCoG states and their computational costs. The table specifies whether each operation updates the log-weights or Gaussian parameters and whether it changes the number of Gaussian terms.
Bosonic operations transform the Gaussian parameters according to the conventional LCoG formalism. Qubit operations leave the Gaussian parameters unchanged, while general qubit operations may increase the number of terms by mixing each qubit block. As shown in Sec.~\ref{sec:4.D}, the state-dependent Gaussian channels preserve the number of Gaussian terms.

\subsection{\label{sec:4.B} Bosonic operations}
Gaussian channels characterized by $\bm X$ and $\bm Y$ transform the covariance matrices and mean vectors in each qubit block:
\begin{equation}
    \begin{aligned}
        \bm\mu_{jkl}&\rightarrow \bm X \bm\mu_{jkl}\\
        \bm{\sigma}_{jkl}&\rightarrow \bm X \bm{\sigma}_{jkl}\bm X^T +\bm Y,
    \end{aligned}
\end{equation}
as given in Eq.~(\ref{eq:gaussian_channel_update}). The log-weights and indices remain unchanged, and thus the channel preserves the number of Gaussian terms. Updating each covariance matrix requires $\mathcal O(M^3)$ operations for each Gaussian term, giving total computational cost of $\mathcal O(N_L M^3)$.

A partial Gaussian measurement also acts independently on each Gaussian term and leaves both the indices and the number of terms unchanged. Suppose that the bosonic subsystem $B_1$ is measured and that $B_2$ denotes the unmeasured subsystem. Applying Eq.~(\ref{eq:partial_gaussian_measurment_update}) to each term gives
\begin{equation}
    \begin{aligned}
        \bm{\sigma}_{\bm j\bm kl}^{B_2}&\mapsto \bm{\sigma}_{\bm j\bm kl}^{B_2}-\bm{\sigma}_{\bm j\bm kl}^{B_1B_2}\left(\bm{\sigma}_{\bm j\bm kl}^{B_1}+\bm{\sigma}_{\hat{\mathcal M}}\right)^{-1}\left(\bm{\sigma}_{\bm j\bm kl}^{B_1B_2}\right)^T\\
        \bm \mu_{\bm j\bm kl}^{B_2}(\mathbf x)&\mapsto \bm \mu_{\bm j\bm kl}^{B_2}+\bm{\sigma}_{\bm j\bm kl}^{B_1B_2}\left(\bm{\sigma}_{\bm j\bm kl}^{B_1}+\bm{\sigma}_{\hat{\mathcal M}}\right)^{-1}\left(\mathbf x-\bm \mu_{\bm j\bm kl}^{B_1}\right),
    \end{aligned}
\end{equation}
The measurement updates the log-weight of each term according to its outcome probability, followed by normalization of the conditional state. The matrix inversion and multiplication require $\mathcal{O}(M^3)$ operations per Gaussian term, and the total computational cost is thus $\mathcal{O}(N_LM^3)$.

For PNRD, the measurement operator is approximated by coherent-state decomposition. When the representation contains $N_\text{PNRD}$ Gaussian terms, each input Gaussian term produces at most $N_\text{PNRD}$ terms in the remaining state.
Therefore, the number of output terms is bounded by $N_\text{PNRD}N_L$, and the computational cost scales as $\mathcal O(N_\text{PNRD} N_L M^3)$.

\subsection{\label{sec:4.C} Qubit operations}
\subsubsection{\label{sec:4.C.1} Qubit gates}
We first consider an $N$-qubit unitary gate $\hat U_A$ acting on the subsystem $A$.
Suppose $u_{\bm i\bm j}$ denotes the $(\bm i,\bm j)$ element of the unitary matrix $\hat U_A$. The bosonic operator in the $(\bar{\bm i},\bar{\bm l})$ block of the output state is given by
\begin{align}
    _{A}\bra{\bar{\bm i}}\hat U_{A}\hat \rho_{AB}\hat U_{A}^\dagger\ket{\bar{\bm l}}_{A}=\sum_{\bm j,\bm k\in\{0,1\}^N} u_{\bm i\bm j}u_{\bm l \bm k}^*\rho_{\bm j\bm k}.
\end{align}
From the above equation, we find that an $N$-qubit unitary recombines the Gaussian functions among qubit blocks without changing their mean vectors and the covariance matrices. Then, each input Gaussian term can contribute to at most $4^N$ output blocks. Consequently, the computational cost is $\mathcal O(4^N N_L)$, and the number of output terms is at most $4^N N_L$. 
Note that this scaling corresponds to the worst case. For example, monomial gates, which only permute the qubit blocks and rotate phase, preserve the number of Gaussians, and their computational cost is $\mathcal O(N_L)$.

More generally, suppose that the gate acts only on an $n$-qubit subsystem $A_1$, while $A_2$ contains the remaining qubits. The full operation is $\hat U_{A_1}\otimes \hat{\mathbb I}_{A_2}$, and each Gaussian term contributes to at most $4^n$ output blocks. Therefore, the computational cost scales as $\mathcal O(4^nN_L)$.

\subsubsection{\label{sec:4.C.2} Qubit quantum channels}
Instead of a unitary gate, we consider a quantum channel $\mathcal E_{A}$ acting on the subsystem $A$. Its action on the hybrid state can be written as
\begin{align}
    \left(\mathcal E_{A}\otimes \mathbb I_{B}\right)\left(\hat \rho_{AB}\right)=\sum_{\bm j,\bm k}\mathcal E_{A}\left(\ket{\bar{\bm j}}\bra{\bar{\bm k}}\right)\otimes\hat\rho_{\bm j\bm k}.
\end{align}
Using a Kraus representation 
\begin{equation}
    \mathcal E_{A}(\hat \rho_{AB})=\sum_i \hat K_i \hat \rho_{AB} \hat K_i^\dagger,
\end{equation}
the channel can be evaluated by the multiplication of each Kraus operator. If Kraus operators act on all $N$ qubits, each input Gaussian term can contribute to at most $4^N$ output blocks for each $i$.
Therefore, the computational cost is $\mathcal O(4^N N_KN_L)$, and the number of output Gaussians is at most $4^N N_KN_L$. 
Typical error channels such as bit-flip and phase-flip channels reduce the computational cost to $\mathcal{O}(N_KN_L)$ and the number of output terms to at most $N_KN_L$ since each Kraus operator is monomial.

If the channel acts only on an $n$-qubit subsystem $A_1$, the full channel is $\mathcal E_{A_1}\otimes \mathbb I_{A_2}$, and the computational cost becomes $\mathcal O(4^n N_K N_L)$.

\subsubsection{\label{sec:4.C.3} Measurements}
We first consider a computational-basis measurement on an $n$-qubit subsystem $A_1$. Conditioned on the outcome $\bar{\bm j}^\prime$, the unnormalized state of the remaining subsystem $A_2 B$ is given by
\begin{align}
    \label{eq:qubit_measurement}
    _{A_1}\bra{\bar{\bm j}^\prime} \hat \rho_{AB}\ket{\bar{\bm j}^\prime}_{A_1}=\hat\rho_{A_2B}^{\bm j^\prime \bm j^\prime}.
\end{align}
The probability of obtaining this outcome is
\begin{align}
    p\left(\bar{\bm j^\prime}\right)=\Tr\left[\hat\rho_{A_2B}^{\bm j^\prime \bm j^\prime}\right],
\end{align}
and the normalized conditional state is $\hat\rho_{A_2B}^{\bm j^\prime \bm j^\prime}/p\left(\bar{\bm j^\prime}\right)$. In the hybrid LCoG representation, this measurement is just selecting the corresponding Gaussian terms and normalizing their weights. Therefore, its computational cost is bounded by $\mathcal O(N_L)$.

A projective measurement in an arbitrary orthonormal basis can be implemented by a unitary gate followed by a computational-basis measurement. Since the unitary gate dominates the computational cost, it scales as $\mathcal O(4^N N_L)$.

\subsection{\label{sec:4.D} Controlled Gaussian operations}
In this section, we derive how controlled Gaussian operations transform the log-weights, complex mean vectors, and complex covariance matrices $\{d_{\bm j \bm kl},\bm \mu_{\bm j\bm k l},\bm{\sigma}_{\bm j\bm k l}\}$ of a hybrid LCoG state. Figure~\ref{fig:controlled-gaussian-operations} shows the operations considered here, and Table~\ref{tab:controlled_gaussian_update} summarizes transformation rules for the operations.
In Secs.~\ref{sec:4.D.1}-\ref{sec:4.D.3}, we consider three single-mode Gaussian operations: displacement, phase rotation, and squeezing. In Sec.~\ref{sec:4.D.4}, we introduce the controlled beam splitter operations, and Sec.~\ref{sec:4.D.5} extends the result to the general multi-controlled multi-mode Gaussian operations. Finally, we describe a controlled Gaussian channel by incorporating noise in Sec.~\ref{sec:4.D.6}. The detailed derivations are in Appendix~\ref{appendix:Weyl_transform}.

\begin{figure}[t]
\centering

\resizebox{\columnwidth}{!}{%
\begin{tabular}{ccc}

\panel{(a)}{
\begin{quantikz}[column sep=0.38cm, row sep=0.30cm]
\lstick{$A$} & \ctrl{1} & \qw \\
\lstick{$B$} & \gate{\hat D(\alpha)} & \qw
\end{quantikz}
}
&
\panel{(b)}{
\begin{quantikz}[column sep=0.38cm, row sep=0.30cm]
\lstick{$A$} & \ctrl{1} & \qw \\
\lstick{$B$} & \gate{\hat R(\theta)} & \qw
\end{quantikz}
}
&
\panel{(c)}{
\begin{quantikz}[column sep=0.38cm, row sep=0.30cm]
\lstick{$A$} & \ctrl{1} & \qw \\
\lstick{$B$} & \gate{\hat S(r)} & \qw
\end{quantikz}
}

\\[5mm]

\panel{(d)}{
\begin{quantikz}[column sep=0.38cm, row sep=0.30cm]
\lstick{$A$}   & \ctrl{1} & \qw \\
\lstick{$B_1$} & \gate[wires=2]{\hat B(\varphi)} & \qw \\
\lstick{$B_2$} &                         & \qw
\end{quantikz}
}
&
\panel{(e)}{
\begin{quantikz}[column sep=0.38cm, row sep=0.30cm]
\lstick{$A$} & \qwbundle{N} & \ctrl{1} & \qw \\
\lstick{$B$} & \qwbundle{M} &\gate[wires=1]{\hat U_G} & \qw \\
\end{quantikz}
}
&
\phantom{
\panel{(c)}{
\begin{quantikz}[column sep=0.38cm, row sep=0.30cm]
\lstick{$q$} & \ctrl{1} & \qw \\
\lstick{$a$} & \gate{S(r)} & \qw
\end{quantikz}
}
}

\end{tabular}
}

\caption{
Examples of controlled Gaussian operations:
(a)~controlled displacement, (b)~controlled phase rotation, (c)~controlled squeezing, (d)~controlled beam splitter, and
(e)~multi-controlled multi-mode Gaussian gates.
}
\label{fig:controlled-gaussian-operations}

\end{figure}
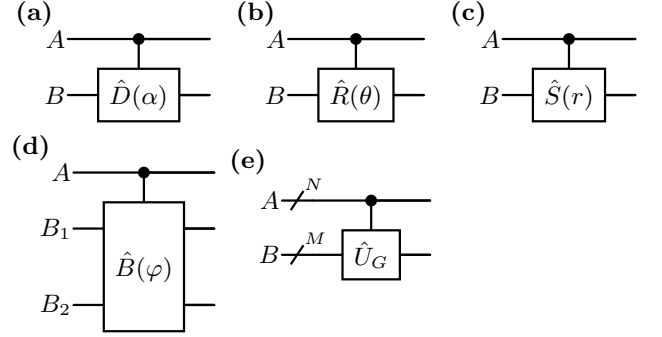
\begin{table*}[htbp]
\caption{Generic transformation rules for a Gaussian term under left and right Gaussian unitaries. Here, $\hat U_L$ and $\hat U_R$ denote the Gaussian unitaries acting from the left and right, respectively. The table gives the resulting changes in the log-weight $d$, mean vector $\bm \mu$, and covariance matrix $\bm \sigma$ with respect to the controlled displacement, phase rotation, squeezing, and beam splitter.}
\label{tab:controlled_gaussian_update}
\scriptsize
\setlength{\tabcolsep}{1pt}
\renewcommand{\arraystretch}{2}
\begin{ruledtabular}
\begin{tabular}{
>{\centering\arraybackslash}p{0.105\textwidth}|
>{\centering\arraybackslash}p{0.260\textwidth}
>{\centering\arraybackslash}p{0.200\textwidth}
>{\centering\arraybackslash}p{0.130\textwidth}
>{\centering\arraybackslash}p{0.255\textwidth}
}
$\hat U_{L/R}$ & Joint operation/vector & $d_{\text{in}}-d_{\text{out}}$ & $\mu_{\text{out}}$ & $\bm{\sigma}_{\text{out}}$ \\[0.4em]\hline
$\hat D\!\left(\frac{\bm\alpha_{L/R}}{\sqrt{2}}\right)$
&
$\mathcal{A}_{LR}=\Pi_0(2\bm{\sigma}_{\text{in}})\bm\alpha_L+\Pi_1(2\bm{\sigma}_{\text{in}})\bm\alpha_R$
&
$\frac{i}{2}(\alpha_L-\alpha_R)^T\Omega(\mathcal{A}_{LR}+2\mu_{\text{in}})$
&
$\mu_{\text{in}}+\mathcal{A}_{LR}$
&
$\bm{\sigma}_{\text{in}}$
\\[0.5em]\hline

$\hat R(\theta_{L/R})$
&
$\mathcal{R}_{LR}(\bm{\sigma})=\Pi_0(\bm{\sigma})\mathsf{R}_L^T+\Pi_1(\bm{\sigma})\mathsf{R}_R^T$
&
$\frac{i}{2}\mu_{\text{in}}^T(\mathsf{R}_L^T-\mathsf{R}_R^T)\Omega\mu_{\text{out}}-\mathcal{N}$
&
$\mathcal{R}_{LR}^{-1}(2\bm{\sigma}_{\text{in}})\mu_{\text{in}}$
&
$\mathcal{R}_{LR}^{-1}(2\bm{\sigma}_{\text{in}})\bm{\sigma}_{\text{in}}\mathcal{R}_{LR}(\bm{\sigma}_{\text{in}}^{-1}/2)$
\\[0.4em]\hline

\vspace{-8pt}$\hat S(r_{L/R})$
&
\vspace{-8pt}
$\begin{gathered}
\mathcal{S}_{LR}(\bm{\sigma})=\Pi_0(\bm{\sigma})\mathsf{S}_L^{-1}+\Pi_1(\bm{\sigma})\mathsf{S}_R^{-1}\\
\mathcal{S}_{LR}^{[T]}(\bm{\sigma})=\Pi_0(\bm{\sigma})\mathsf{S}_L^T+\Pi_1(\bm{\sigma})\mathsf{S}_R^T
\end{gathered}$
&
\vspace{-8pt}
$\begin{gathered}
\frac{i}{2}\mu_{\text{in}}^T(\mathsf{S}_L^T-\mathsf{S}_R^T)\Omega\mu_{\text{out}}-\mathcal{N}_\text{sq}
\end{gathered}$
&
\vspace{-8pt}
$\mathcal{S}_{LR}^{-1}(2\bm{\sigma}_{\text{in}})\mu_{\text{in}}$
&
\vspace{-8pt}
$\mathcal{S}_{LR}^{-1}(2\bm{\sigma}_{\text{in}})\bm{\sigma}_{\text{in}}\mathcal{S}_{LR}^{[T]}(\bm{\sigma}_{\text{in}}^{-1}/2)$
\\[1.5em]\hline

$\hat B(\varphi_{L/R})$
&
$\mathcal{B}_{LR}(\bm{\sigma})=\Pi_0(\bm{\sigma})\mathsf{B}_L^T+\Pi_1(\bm{\sigma})\mathsf{B}_R^T$
&
$\frac{i}{2}\mu_{\text{in}}^T(\mathsf{B}_L^T-\mathsf{B}_R^T)\Omega\mu_{\text{out}}-\mathcal{N}$
&
$\mathcal{B}_{LR}^{-1}(2\bm{\sigma}_{\text{in}})\mu_{\text{in}}$
&
$\mathcal{B}_{LR}^{-1}(2\bm{\sigma}_{\text{in}})\bm{\sigma}_{\text{in}}\mathcal{B}_{LR}(\bm{\sigma}_{\text{in}}^{-1}/2)$
\end{tabular}
\end{ruledtabular}
\end{table*}

\subsubsection{\label{sec:4.D.1} Controlled displacement}
A controlled displacement gate is defined by 
\begin{align}
    \widehat{\mathsf{CD}}(\bm \alpha)=\ket{\bar 0}\bra{\bar 0}\otimes \hat D\left(\frac{\bm \alpha_0}{\sqrt 2}\right)+\ket{\bar 1}\bra{\bar 1}\otimes \hat D\left(\frac{\bm \alpha_1}{\sqrt2}\right),
\end{align}
where $\bm \alpha=(\bm \alpha_0, \bm \alpha_1)$ represents a pair of displacement amplitudes, and $\hat D(\bm \alpha)$ denotes the Weyl operator defined in Eq.~(\ref{eq:Weyl_operator}).

To derive the transformation rule, we consider the single-qubit and single-mode state as given in Eq.~(\ref{eq:single-qubit-oscillator-state}).
The controlled displacement transforms it as
\begin{align}
    &\widehat{\mathsf{CD}}(\bm \alpha)\hat \rho_\text{q,b}\widehat{\mathsf{CD}}^\dagger(\bm \alpha)\nonumber\\
    =&\sum_{j,k\in\{0,1\}}\ket{\bar j}\bra{\bar k}\otimes \hat D\left(\frac{\bm \alpha_j}{\sqrt 2}\right)\hat\rho_{jk}\hat D^\dagger\left(\frac{\bm \alpha_k}{\sqrt 2}\right).
\end{align}
For a diagonal block, $j=k$, the same displacement acts from the left and right. Its transformation thus follows the standard Gaussian transform in Sec.~\ref{sec:4.B}. For off-diagonal block, $j\neq k$, the two displacements are generally different:
\begin{equation}
    \hat \rho_{jk}\mapsto \hat D\left(\frac{\bm \alpha_j}{\sqrt 2}\right)\hat\rho_{jk}\hat D^\dagger\left(\frac{\bm \alpha_k}{\sqrt 2}\right)
\end{equation}
This action cannot be described by the standard Gaussian transformation. Therefore, we evaluate it through the Weyl transform. 
By expanding the $(j,k)$ block into LCoG form,
\begin{equation}
    \begin{aligned}
        &\hat \rho_{jk}=\sum_{l=0}^{L_{jk}-1}\hat \varrho_{jkl},\qquad \text{with}\\
        &W_{\hat\varrho_{jkl}}(\bm q)=e^{d_{jkl}}G_{\bm \mu_{jkl},\bm \sigma_{jkl}}(\bm q),
    \end{aligned}
\end{equation}
we can treat each term independently:
\begin{align}
    \label{eq:displaced_off_diagonal_block}
    \hat D\left(\frac{\bm \alpha_j}{\sqrt 2}\right)\rho_{jk}\hat D^\dagger\left(\frac{\bm \alpha_k}{\sqrt 2}\right)=\sum_{l=0}^{L_{jk}-1}\hat\varrho_{jkl}^\text{out},
\end{align}
where
\begin{equation}
    \hat\varrho_{jkl}^\text{out}=\hat D\left(\frac{\bm \alpha_j}{\sqrt 2}\right)\varrho_{jkl}\hat D^\dagger\left(\frac{\bm \alpha_k}{\sqrt 2}\right).
\end{equation}
The Weyl transform of the $l$th term is given by
\begin{equation}
    \begin{aligned}
        \label{eq:Weyl_transform_of_displaced_off_diagonal_block}
        &W_{\hat\varrho_{jkl}^\text{out}}(\bm q)\\
        =&\frac{1}{2\pi^2}\int d^2\bm \zeta\ \Tr[\hat D^\dagger\left(\frac{\bm \alpha_k}{\sqrt 2}\right)\hat\Theta_{\bm q,\bm \zeta}\hat D\left(\frac{\bm \alpha_j}{\sqrt 2}\right)\hat\varrho_{jkl}]e^{i\bm q^T \bm \Omega\bm \zeta}.
    \end{aligned}
\end{equation}
The operator $\hat D^\dagger\left(\frac{\bm \alpha_k}{\sqrt 2}\right)\hat\Theta_{\bm q,\bm \zeta}\hat D\left(\frac{\bm \alpha_j}{\sqrt 2}\right)$ remains an outer product of coherent states and therefore has a complex Gaussian Weyl transform, as given in Eq.~(\ref{eq:params_outer_product_of_coherent_state}). Since Weyl transform of $\varrho_{jkl}$ is also Gaussian, the trace in Eq.~(\ref{eq:Weyl_transform_of_displaced_off_diagonal_block}) reduces to a Gaussian integral. By evaluating this integral, we find the following transformation of the mean vector, covariance matrix, and log-weight:
\begin{subequations}
    \label{eq:updating_rule_cd}
    \begin{align}
        \bm \mu_{jkl}^\text{out}&=\bm\mu_{jkl}+\bm{\mathcal{A}}_{jkl},\\
        \bm{\sigma}_{jkl}^\text{out}&=\bm{\sigma}_{jkl},\\
        d_{jkl}^\text{out}&= d_{jkl}-\frac{i}{2}(\bm \alpha_j-\bm \alpha_k)^T\Omega\left(\bm{\mathcal{A}}_{jkl}+2\bm\mu_{jkl}\right),
    \end{align}
\end{subequations}
where $\bm{\mathcal A}_{jkl}$ denotes the joint displacement vector defined by
\begin{equation}
    \label{eq:joint_displacement}
    \bm{\mathcal A}_{jkl} =\Pi_0{(2\bm{\sigma}_{jkl})}\bm \alpha_j+\Pi_1{(2\bm{\sigma}_{jkl})}\bm \alpha_k, 
\end{equation}
and the matrix $\Pi_s(\bm{\sigma})$ represents the left and right action:
\begin{equation}
    \Pi_s{(\bm{\sigma})}=\frac{1}{2}\left[\mathbb I +(-1)^si \bm{\sigma}\Omega\right].
\end{equation}
The two terms in Eq.~(\ref{eq:joint_displacement}) encode the displacements applied from the left and right, respectively. For $j=k$, Eq.~(\ref{eq:updating_rule_cd}) reduces to the standard Gaussian transformation.

\subsubsection{\label{sec:4.D.2} Controlled phase rotation}
A controlled phase-rotation gate is defined by
\begin{align}
    \widehat{\mathsf{CR}}(\bm \theta)=\ket{\bar 0}\bra{\bar 0}\otimes \hat R(\theta_0)+\ket{\bar 1}\bra{\bar 1}\otimes \hat R(\theta_1),
\end{align}
where $\bm \theta=(\theta_0,\theta_1)$ denotes a pair of rotation angles, and $\hat R(\theta)=\exp(-i\theta \hat a^\dagger \hat a)$ represents the phase rotation operator with the creation and annihilation operators, $\hat a^\dagger$ and $\hat a$, respectively.

We again consider the single-qubit and single-mode state in Eq.~(\ref{eq:single-qubit-oscillator-state}). The controlled phase rotation transforms the state as
\begin{align}
    &\widehat{\mathsf{CR}}(\bm \theta)\hat \rho_\text{q,b}\widehat{\mathsf{CR}}^\dagger(\bm \theta)\nonumber\\
    =&\sum_{j,k\in\{0,1\}}\ket{\bar j}\bra{\bar k}\otimes \hat R(\theta_j)\hat\rho_{jk}\hat R^\dagger(\theta_k).
\end{align}
Following the derivation for the controlled displacement, the Weyl transform of the $l$th Gaussian term in the $(j,k)$ block is
\begin{equation}
    \begin{aligned}
        \label{eq:Weyl_transform_of_phase_rotated_off_diagonal_block}
        &W_{\hat\varrho_{jkl}^\text{out}}(\bm q)\\
        =&\frac{1}{2\pi^2}\int d^2\bm \zeta\ \Tr[\hat R^\dagger(\theta_k)\hat\Theta_{\bm q,\bm \zeta}\hat R(\theta_j)\hat\varrho_{jkl}]e^{i\bm q^T \bm \Omega\bm \zeta}.
    \end{aligned}
\end{equation}
A phase rotation maps a coherent state to another coherent state. Consequently, $\hat R^\dagger(\theta_k)\hat\Theta_{\bm q,\bm \zeta}\hat R(\theta_j)$ remains an outer product of two different coherent states.
Therefore, the trace in Eq. (\ref{eq:Weyl_transform_of_phase_rotated_off_diagonal_block}) reduces to a Gaussian integral. Evaluating it gives
\begin{subequations}
    \label{eq:updating_rule_cr}
    \begin{align}
        \bm \mu_{jkl}^\text{out}=&\mathcal{R}_{jk}^{-1}\left(2\bm{\sigma}_{jkl}\right)\bm \mu_{jkl},\\
        \bm{\sigma}_{jkl}^\text{out}=&\mathcal{R}_{jk}^{-1}\left(2\bm{\sigma}_{jkl}\right)\bm{\sigma}_{jkl}\mathcal{R}_{jk}\left(\bm{\sigma}_{jkl}^{-1}/2\right),\\
        d_{jkl}^\text{out}=& d_{jkl}-\frac{i}{2}\bm \mu_{jkl}^T\left(\mathsf{R}_j^T-\mathsf{R}_k^T\right)\Omega\bm \mu_{jkl}^\text{out}+\mathcal{N},\\
        \mathcal N=&\ln \sqrt \frac{\det \bm{\sigma}_{jkl}^\text{out}+\mathbb I/2}{\det \bm{\sigma}_{jkl}+\mathbb I/2},
    \end{align}
\end{subequations}
where $\mathcal R_{jk}(\bm{\sigma})$ is the joint phase-rotation matrix,
\begin{equation}
    \label{eq:joint_phase_rotation_matrix}
    \mathcal R_{jk}(\bm{\sigma})=\Pi_0(\bm{\sigma})\mathsf{R}_j^T+\Pi_1(\bm{\sigma})\mathsf{R}_k^T,
\end{equation}
with the symplectic matrices $\mathsf{R}_j$ and $\mathsf{R}_k$ corresponding to the phase-rotation operators $\hat R(\theta_j)$ and $\hat R(\theta_k)$, respectively. The logarithmic determinant $\mathcal N$ accounts for the change in the normalization of the Gaussian function. For $j=k$, Eq.~(\ref{eq:joint_phase_rotation_matrix}) gives $\mathcal R_{jj}(\bm \sigma)=\mathsf{R}_j^T$, and the transformation rule reduces to the standard Gaussian transformation.

\subsubsection{\label{sec:4.D.3} Controlled squeezing}
A controlled squeezing gate is defined by
\begin{align}
    \widehat{\mathsf{CS}}(\bm r)=\ket{\bar 0}\bra{\bar 0}\otimes \hat S(r_0)+\ket{\bar 1}\bra{\bar 1}\otimes \hat S(r_1),
\end{align}
where $\bm r=(r_0,r_1)$ denotes a pair of squeezing parameters, and $\hat S(r)=\exp[-r({a^\dagger}^2-a^2)/2]$ denotes a squeezing operator.

The controlled squeezing transforms the input state as
\begin{align}
    &\widehat{\mathsf{CS}}(\bm r)\hat \rho_\text{q,b}\widehat{\mathsf{CS}}^\dagger(\bm r)\nonumber\\
    =&\sum_{j,k\in\{0,1\}}\ket{\bar j}\bra{\bar k}\otimes \hat S(r_j)\hat \rho_{jk}\hat S^\dagger(r_k).
\end{align}
Following the derivation for the controlled displacement, the Weyl transform of the $l$th Gaussian term in the $(j,k)$ block is
\begin{equation}
    \begin{aligned}
        \label{eq:Weyl_transform_of_squeezed_off_diagonal_block}
        &W_{\hat \varrho_{jkl}^\text{out}}(\bm q)\\
        =&\frac{1}{2\pi^2}\int d^2\bm \zeta\ \Tr[\hat S^\dagger(r_k)\hat\Theta_{\bm q,\bm \zeta}\hat S(r_j)\hat \varrho_{jkl}]e^{i\bm q^T \bm \Omega\bm \zeta}.
    \end{aligned}
\end{equation}
Unlike a phase rotation, squeezing maps coherent states to displaced squeezed states. Nevertheless, $\hat S^\dagger(r_k)\hat\Theta_{\bm q,\bm \zeta}\hat S(r_j)$ is still an outer product of pure Gaussian states, and therefore its Weyl transform is complex Gaussian, as derived in Appendix~\ref{appendix:controlled_squeezing},
\begin{equation}
    W_{\hat S^\dagger(r_k)\hat \Theta_{\bm q,\bm \zeta}\hat S(r_j)}(\bm q)=e^{\delta_{jk}}G_{\bm{\nu}_{jk},\bm \varsigma_{jk}}(\bm q).
\end{equation}
Substituting this result into Eq.~(\ref{eq:Weyl_transform_of_squeezed_off_diagonal_block}) reduces the remaining calculation to a Gaussian integral and yields
\begin{subequations}
    \label{eq:updating_rule_cs}
    \begin{align}
        \bm \mu_{jkl}^\text{out}=&\mathcal{S}_{jk}^{-1}(2\bm{\sigma}_{jkl})\bm \mu_{jkl},\\
        \bm{\sigma}_{jkl}^\text{out}=&\mathcal{S}_{jk}^{-1}(2\bm{\sigma}_{jkl})\bm{\sigma}_{jkl}\mathcal{S}_{jk}^{[T]}(\bm{\sigma}_{jkl}^{-1}/2),\\
        d_{jkl}^\text{out}=& d_{jkl}-\frac{i}{2}\bm \mu_{jkl}^T\left(\mathsf{S}_j^T-\mathsf{S}_k^T\right)\Omega\bm \mu_{jkl}^\text{out}+\mathcal N_\text{sq}\\
        \mathcal N_\text{sq}=&\ln \sqrt \frac{\det \bm{\sigma}_{jkl}^\text{out}+\mathbb I/2}{\cosh(r_j-r_k)\det (\bm{\sigma}_{jkl}+\bm \varsigma_{jk})}.
    \end{align}
\end{subequations}
Here, $\varsigma_{jk}$ is the complex covariance matrix associated with the outer product of the two squeezed vacuum states, 
\begin{equation}
    \bm \varsigma=\frac{(1+\lambda_j\lambda_k)\mathbb I+2Z\left(\lambda_k\Pi_1+\lambda_j\Pi_0\right)}{2(1-\lambda_j\lambda_k)},
\end{equation}
where $\lambda_j=\tanh r_j$, $\lambda_k=\tanh r_k$, and $Z=\text{diag}(1,-1)$. 
The two joint squeezing matrices are defined by
\begin{subequations}
    \label{eq:joint_squeezing_jk}
    \begin{align}
        \mathcal S_{jk}(\bm{\sigma})&=\Pi_0(\bm{\sigma})\mathsf{S}_j^{-1}+\Pi_1(\bm{\sigma})\mathsf{S}_k^{-1},\\
        \mathcal S_{jk}^{[T]}(\bm{\sigma})&=\Pi_0(\bm{\sigma})\mathsf{S}_j^{T}+\Pi_1(\bm{\sigma})\mathsf{S}_k^{T}
    \end{align}
\end{subequations}
where the symplectic matrices $\mathsf{S}_j$ and $\mathsf{S}_k$ correspond to the squeezing operators $\hat S(r_j)$ and $\hat S(r_k)$, respectively. For $j=k$, Eq.~(\ref{eq:updating_rule_cs}) reproduces the standard squeezing transformation.

\subsubsection{\label{sec:4.D.4} Controlled beam splitter}
A controlled beam splitter is defined as
\begin{align}
    \widehat{\mathsf{CB}}(\bm\varphi)=\ket{\bar 0}\bra{\bar 0}\otimes \hat{B}(\varphi_0)+\ket{\bar 1}\bra{\bar 1}\otimes \hat{B}(\varphi_1),
\end{align}
where $\bm \varphi=(\varphi_0,\varphi_1)$ denotes a pair of beam splitter angles, and $\hat B(\varphi)=\exp[\varphi(\hat a_{1}\hat a_{2}^\dagger-\hat a_{1}^\dagger \hat a_{2})]$ is the beam splitter operator. 

We consider a state $\hat \rho_\text{AB}$ composed of a single qubit and a two-mode bosonic subsystem $B$. The controlled beam splitter transforms the state as
\begin{align}
    &\widehat{\mathsf{CB}}(\bm \varphi)\hat \rho_\text{AB}\widehat{\mathsf{CB}}^\dagger(\bm \varphi)\nonumber\\
    =&\sum_{j,k\in\{0,1\}}\ket{\bar j}\bra{\bar k}\otimes \hat B(\varphi_j)\hat \rho_{jk}\hat B^\dagger(\varphi_k).
\end{align}

Following the derivation for the controlled displacement, the Weyl transform of the $l$th Gaussian term in the $(j,k)$ block is
\begin{align}
    \label{eq:Weyl_transform_of_off_diagonal_bs}
    &W_{\hat \varrho_{jkl}^\text{out}}(\bm q)\nonumber\\
    =&\frac{1}{(2\pi^2)^2}\int d^4\bm \zeta\ \Tr[\hat B^\dagger(\varphi_k)\hat\Theta_{\bm q,\bm \zeta}\hat B(\varphi_j)\hat \varrho_{jkl}]e^{i\bm q^T \bm \Omega\bm \zeta}.
\end{align}
A beam splitter maps a coherent state to another coherent state. Therefore, $\hat B^\dagger(\varphi_k)\hat\Theta_{\bm q,\bm \zeta}\hat B(\varphi_j)$ remains an outer product of multi-mode coherent states and has a complex Gaussian Weyl transform. The trace in Eq.~(\ref{eq:Weyl_transform_of_off_diagonal_bs}) thus reduces to a Gaussian integral, as in the controlled phase-rotation case. Evaluating this integral gives
\begin{subequations}
    \label{eq:updating_rule_cb}
    \begin{align}
        \bm \mu_{jkl}^\text{out}=&\mathcal{B}_{jk}^{-1}(2\bm{\sigma}_{jkl})\bm \mu_{jkl},\\
        \bm{\sigma}_{jkl}^\text{out}=&\mathcal{B}_{jk}^{-1}(2\bm{\sigma}_{jkl})\bm{\sigma}_{jkl}\mathcal{B}_{jk}(\bm{\sigma}_{jkl}^{-1}/2),\\
        d_{jkl}^\text{out}=& d_{jkl}-\frac{i}{2}\bm \mu_{jkl}^T\left(\mathsf{B}_j^T-\mathsf{B}_k^T\right)\Omega\bm \mu_{jkl}^\text{out}+\mathcal N\\
        \mathcal{N}=&\ln \sqrt \frac{\det \bm{\sigma}_{jkl}^\text{out}+\mathbb I/2}{\det \bm{\sigma}_{jkl}+\mathbb I/2},
    \end{align}
\end{subequations}
where $\mathcal B_{jk}(\bm{\sigma})$ is the joint beam splitter matrix,
\begin{equation}
    \mathcal B_{jk}(\bm{\sigma})=\Pi_0(\bm{\sigma})\mathsf{B}_j^T+\Pi_1(\bm{\sigma})\mathsf{B}_k^T,
\end{equation}
with the symplectic matrices $\mathsf{B}_j$ and $\mathsf{B}_k$ corresponding to the beam splitter operators $\hat B(\varphi_j)$ and $\hat B(\varphi_k)$, respectively. 

An arbitrary passive multi-mode Gaussian unitary can be decomposed into beam splitters and phase rotations. We denote such a multi-mode interferometer by $\hat V$. Since both passive operations preserve multi-mode coherent states, the above derivation can be used without modification. The resulting transformation has the same form:
\begin{subequations}
    \label{eq:updating_rule_cv}
    \begin{align}
        \bm \mu_{jkl}^\text{out}=&\mathcal{V}_{jk}^{-1}(2\bm{\sigma}_{jkl})\bm \mu_{jkl},\\
        \bm{\sigma}_{jkl}^\text{out}=&\mathcal{V}_{jk}^{-1}(2\bm{\sigma}_{jkl})\bm{\sigma}_{jkl}\mathcal{V}_{jk}(\bm{\sigma}_{jkl}^{-1}/2),\\
        d_{jkl}^\text{out}=& d_{jkl}-\frac{i}{2}\bm \mu_{jkl}^T\left(\mathsf{V}_j^T-\mathsf{V}_k^T\right)\Omega\bm \mu_{jkl}^\text{out}+\mathcal N\\
        \mathcal N=&\ln \sqrt \frac{\det \bm{\sigma}_{jkl}^\text{out}+\mathbb I/2}{\det \bm{\sigma}_{jkl}+\mathbb I/2},
    \end{align}
\end{subequations}
where $\mathcal V_{jk}(\bm{\sigma})$ is the joint multi-mode beam splitter matrix,
\begin{equation}
    \mathcal V_{jk}(\bm{\sigma})=\Pi_0(\bm{\sigma})\mathsf{V}_j^T+\Pi_1(\bm{\sigma})\mathsf{V}_k^T,
\end{equation}
with the symplectic matrices $\mathsf{V}_j$ and $\mathsf{V}_k$ corresponding to the multi-mode beam splitter operators $\hat V_j$ and $\hat V_k$, respectively. 

\subsubsection{\label{sec:4.D.5} Multi-controlled Gaussian operations}
So far, we have considered a single control qubit and specific Gaussian operations. In this section, we generalize these results to a subsystem $A$ composed of $N$ control qubits and a subsystem $B$ composed of $M$ modes. The most general controlled Gaussian unitary that is diagonal in the computational basis of subsystem A can be written as
\begin{align}
    \widehat{\mathsf{CG}}_{A B}=\sum_{\bm j\in \{0,1\}^N}\ket{\bar{\bm j}}_{A}\bra{\bar{\bm j}}\otimes \hat U_{\bm j},
\end{align}
where $\hat U_{\bm j}$ denotes an arbitrary Gaussian unitary.
Note that the unitary operator $\hat U_{\bm j}$ may coincide for different $\bm j$, allowing us to describe all possible controlled Gaussian operations.

For a quantum state $\hat \rho_\text{AB}$, the controlled Gaussian unitary gives
\begin{align}
    \widehat{\mathsf{CG}}\hat\rho_{AB}\widehat{\mathsf{CG}}^\dagger=\sum_{\bm j,\bm k\in\{0,1\}^N}\ket{\bar{\bm j}}\bra{\bar{\bm k}}\otimes \hat U_{\bm j}\hat \rho_{\bm j\bm k}\hat U_{\bm k}^\dagger.
\end{align}

Using the Bloch-Messiah decomposition, each $M$-mode Gaussian unitary $\hat U_{\bm j}$ is decomposed as
\begin{align}
    \hat U_{\bm j}=\hat D_M(\bm \alpha_{\bm j})\hat V_{\bm j} \left(\bigotimes_{m=1}^M\hat S_m(r_{m\bm j})\right)\hat V_{\bm j}^\prime,
\end{align}
where $\bm \alpha_{\bm j}$ is the $M$-mode displacement vector, $\hat V_{\bm j}$ and $\hat V_{\bm j}^\prime$ are passive $M$-mode Gaussian unitaries, and $r_{m\bm j}$ is the squeezing parameter applied to the mode $m$.
Therefore, the multi-controlled Gaussian operation $\widehat{\mathsf{CG}}_{AB}$ can be decomposed into a multi-controlled displacement, two multi-controlled passive Gaussian unitaries, and multi-controlled single-mode squeezers.

Since each input Gaussian term is transformed into a single Gaussian term, a multi-qubit-controlled Gaussian unitary preserves the total number $N_L$ of Gaussian terms. For $2M\times 2M$ matrices, the matrix inversions, multiplications, and determinants scale as $\mathcal O(M^3)$. The total computational cost is $\mathcal O(N_LM^3)$.

\subsubsection{\label{sec:4.D.6} Controlled Gaussian channels}
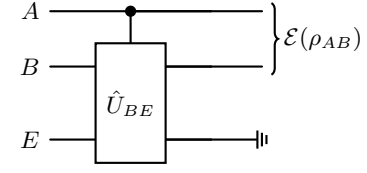
\begin{figure}
    \centering
    \begin{quantikz}[column sep=0.60cm, row sep=0.35cm]
        \lstick{$A$} & \ctrl{1} & \qw & \rstick[wires=2]{$\mathcal E(\rho_{AB})$} \\
        \lstick{$B$} & \gate[wires=2]{\hat U_{BE}} & \qw & \\
        \lstick{$E$} &                  & \qw & \ground{}
    \end{quantikz}
    \caption{Stinespring dilation of a controlled Gaussian channel. A controlled Gaussian unitary $\hat U{BE}$ acts on both the bosonic subsystem $B$ and the environment $E$. The state in the environment is initially in a Gaussian state. Tracing out the environment leads to a Gaussian channel $\mathcal E(\rho_{AB})$.}
    \label{fig:state_dependent_Gaussian_Channel}
\end{figure}
During controlled Gaussian operations, the target bosonic modes may couple to an environment $E$ with different interaction strengths depending on the qubit states. As shown in Fig.~\ref{fig:state_dependent_Gaussian_Channel}, tracing out the environment leads to a controlled Gaussian channel on the hybrid system.

Suppose that a Gaussian state $\hat \rho_E$ is initially prepared in $M_E$-mode environment.
A controlled Gaussian operation over bosonic subsystem $B$ and the environment $E$ is represented by
\begin{align}
    \widehat{\mathsf{CG}}_{ABE}=\sum_{\bm j\in\{0,1\}^N}\ket{\bar{\bm j}}\bra{\bar{\bm j}}\otimes \hat U_{\bm j},
\end{align}
where $\hat U_{\bm j}$ denotes a Gaussian unitary acting on $B$ and $E$.
Using Stinespring dilation, the resulting controlled Gaussian channel is given by
\begin{align}
    \mathcal E_{AB}(\rho_{AB})=\Tr_E\left[\widehat{\mathsf{CG}}_{ABE} \left(\hat\rho_{AB}\otimes\hat \rho_E\right)\widehat{\mathsf{CG}}_{ABE}^\dagger\right].
\end{align}
The channel can be evaluated by applying the transformation rules derived in Sec.~\ref{sec:4.D.5} and subsequently tracing out the environment. Since each input Gaussian term is mapped to a single Gaussian term, the channel preserves the total number $N_L$ of Gaussians. For covariance matrices of $M+M_E$-mode system, the computational cost scales as $\mathcal{O}\left[N_L(M+M_E)^3\right]$.

\begin{figure}[t]
    \centering
    \includegraphics[width=1\linewidth]{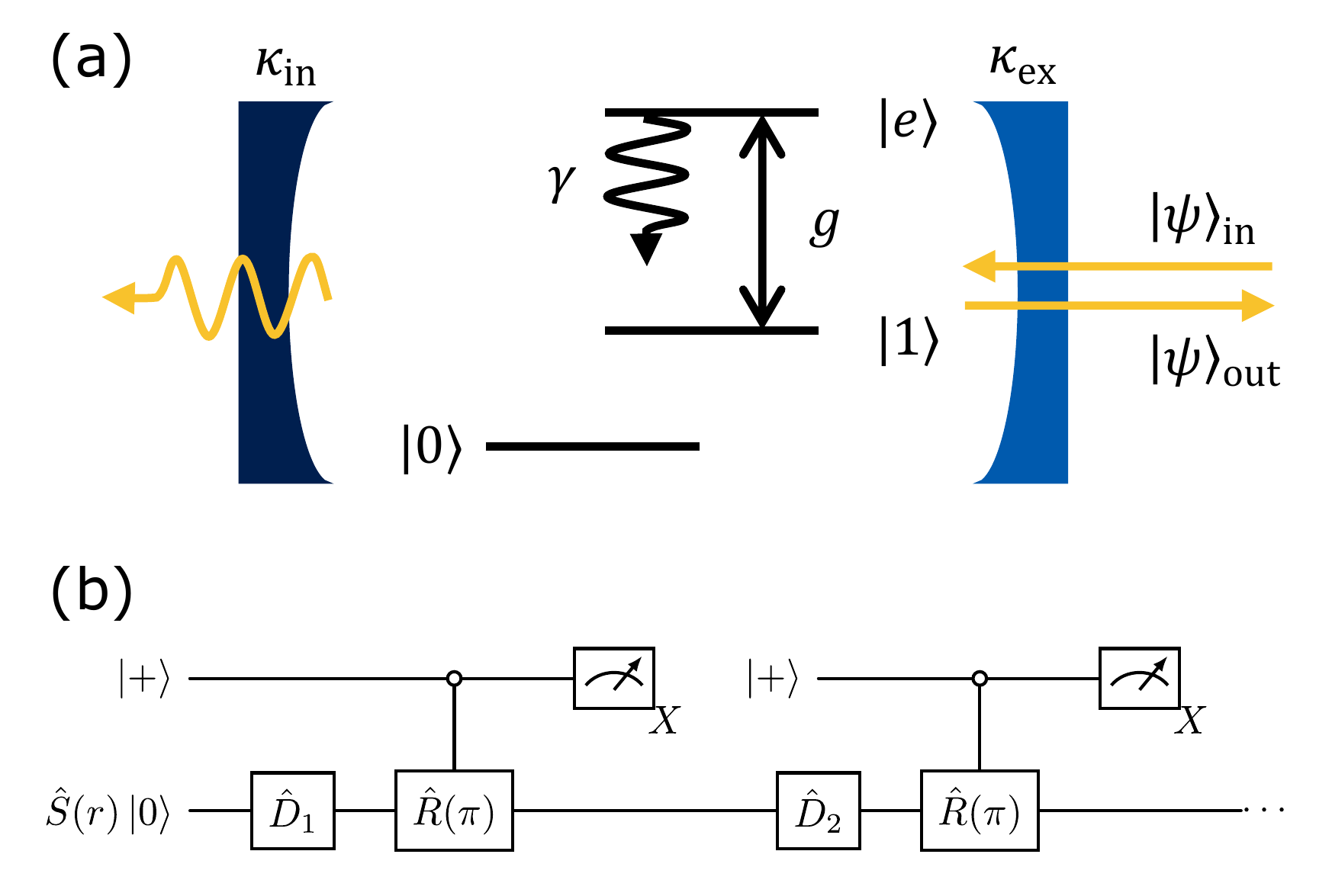}
    \caption{Implementation of the controlled phase rotation gate on the cavity-QED system and the circuit for GKP state generation. (a) Cavity-QED system with atom-cavity coupling $g$, intracavity loss rate $\kappa_\text{in}$, and external coupling rate $\kappa_\text{ex}$. (b) Encoding circuit for a GKP state based on iterative measurements.}
    \label{fig:cavity-QED_setup}
\end{figure}

\section{Example: Cavity-based GKP state generation}

In this section, we consider the approximate GKP state generation from a cavity-QED system~\cite{hastrupProtocolGeneratingOptical2022a} as an example of our hybrid LCoG framework. 
The setup of the cavity-QED system is shown in Fig.~\ref{fig:cavity-QED_setup}(a). The qubit is encoded in two ground states, $\ket0$ and $\ket 1$, of the three-level atom. The transition $\ket{1}\leftrightarrow\ket e$ couples to a one-sided cavity with coupling $g$. The cavity couples to the external propagating field at rate $\kappa_\text{ex}$ and has an internal loss rate $\kappa_\text{in}$. The spontaneous decay rate from the excited state is denoted by $\gamma$. For a monochromatic input in the absence of loss, reflection implements the controlled phase rotation
\begin{equation}
    \widehat{\mathsf{CR}}(\pi)=\ket 0\bra 0 \otimes e^{i\pi\hat n}+\ket 1\bra 1 \otimes\hat{\mathbb I}.
\end{equation}

The approximate GKP states are represented by superposition of displaced squeezed states:
\begin{equation}
    \begin{aligned}
        \ket{\bar 0}_L\propto& \sum_{s}\hat D\left(\bm \beta s\right)\hat S(r)\ket{0},\\
        \ket{\bar 1}_L\propto&\hat D\left(\bm\beta/2\right)\ket{0}_L,
    \end{aligned}
\end{equation}
where $\beta=[\sqrt{2\pi},0]^T$.
The state can be probabilistically generated by repeating the circuit as shown in Fig.~\ref{fig:cavity-QED_setup}(b). In the $n$th encoding block, the photonic state is first displaced by $\hat D_n=\hat D(2^{n-2}\bm \beta)$, and subsequent indirect measurement is performed by the controlled phase-rotation and Pauli-X measurement. For the measurement outcome $\ket{+}\bra{+}$, the $n$th encoding block implements
\begin{align}
    \hat M_{+,n}=\frac{1}{2} \left[\hat{\mathbb I}+\hat R(\pi)\right]\hat D(2^{n-2}\bm \beta).
\end{align}
Therefore, the encoding block doubles the number of terms in the state vector.

\begin{figure}
    \centering
    \includegraphics[width=1\linewidth]{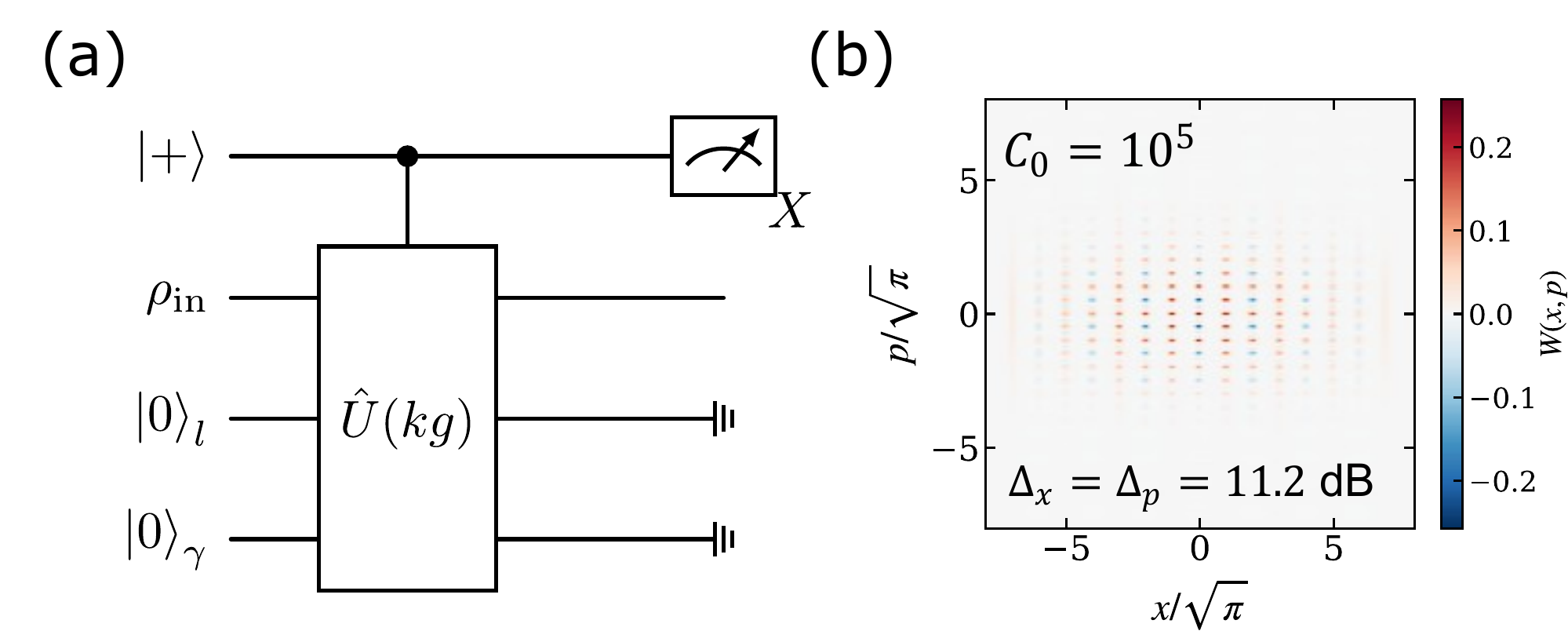}
    \caption{(a) Circuit model of the controlled phase-rotation gate based on the cavity-QED system. The Gaussian unitary $\hat U(kg)$ depends on the qubit state $k$, reflecting state-dependent atom-cavity coupling. Here, $\rho_\text{in}$ denotes the input state, $\ket{0}_l$ denotes the vacuum state at the intracavity loss mode, and $\ket{0}_\gamma$ denotes the vacuum state at the atomic scattering mode. These two loss modes are traced out after the interaction. (b) Generated GKP state with $\Delta_x=\Delta_p=11.2$ dB effective squeezing.}
    \label{fig:cavity_QED_channel}
\end{figure}

Internal cavity loss and atomic scattering introduce errors into the generated GKP state.
The photon is scattered by the atom in $\ket 1$, and the reflection and internal loss also depend on the atomic state. 
Therefore, the cavity-QED system is described by a controlled Gaussian channel, as shown in Fig.~\ref{fig:cavity_QED_channel}(a). The controlled unitary operation is given by
\begin{align}
    \label{eq:controlled_phase_based_on_cavity_QED}
    \hat U_\text{cav}=\ket{0}\bra{0}\otimes \hat U(0)+\ket{1}\bra{1}\otimes \hat U(g).
\end{align}
where the unitary operator $\hat U(kg)$ $(k\in\{0,1\})$ is a three-mode beam splitter acting on the reflected field and two loss modes. Since the unitary operators are passive linear operators, the action can be computed by Eq.~(\ref{eq:updating_rule_cv}) along with the corresponding symplectic matrices $\mathsf{U}(0)$ and $\mathsf{U}(g)$ derived in Appendix~\ref{appendix:cQED_controlled_unitary}:
\begin{widetext}
\begin{equation}
    \label{eq:symplectic_matrices_in_cavity_QED}
    \begin{aligned}
        \mathsf{U}(0)&=
        \begin{bmatrix}
            (1-2\eta)I&-2\sqrt{\eta(1-\eta)}I&0\\
            -2\sqrt{\eta(1-\eta)}I&(-1+2\eta)I&0\\
            0&0&-I
        \end{bmatrix},\\
        \mathsf{U}(g)&=\frac{1}{1+2C}
        \begin{bmatrix}
            (1+2C-2\eta)I&-2\sqrt{\eta(1-\eta)}I&-2\sqrt{2\eta C}\,iY\\
            -2\sqrt{\eta(1-\eta)}I&(-1+2C+2\eta)I &-2\sqrt{2(1-\eta)C}iY\\
            -2\sqrt{2\eta C}iY&-2\sqrt{2(1-\eta)C}iY&(-1+2C)I
        \end{bmatrix},
    \end{aligned}
\end{equation}
\end{widetext}
where $C=g^2/(2\kappa\gamma)$ denotes the cooperativity, $\eta=\kappa_\text{ex}/\kappa$ denotes the escape efficiency, and
\begin{equation}
    \begin{alignedat}{2}
        I&=\begin{bmatrix}
            1 & 0\\
            0 & 1
        \end{bmatrix},
        &\qquad
        Y&=\begin{bmatrix}
            0 & -i\\
            i & 0 
        \end{bmatrix}.
    \end{alignedat}
\end{equation}
Figure~\ref{fig:cavity_QED_channel}(b) shows the generated GKP state after three encoding blocks. Here, we characterize the generated GKP state by effective squeezing~\cite{Duivenvoorden2017}:
\begin{equation}
    \begin{aligned}
        \Delta_x=&\sqrt{\frac{1}{2\pi}\ln\abs{\langle\hat D(\bm \Omega\bm \beta)\rangle}^{-2}},\\
        \Delta_p=&\sqrt{\frac{1}{2\pi}\ln\abs{\langle\hat D(\bm \beta)\rangle}^{-2}}.
    \end{aligned}
\end{equation}
We set the internal cooperativity $C_0=C/(1-\eta)$ to $10^5$ and used optimized internal loss rate $\kappa_\text{in}$ and external coupling rate $\kappa_\text{ex}$ as given in Refs.~\cite{hastrupProtocolGeneratingOptical2022a,kikuraPassiveQuantumInterconnects2025}. The remaining parameters including the input squeezing and the displacement at each encoding block were numerically optimized. 
To produce the envelope of the approximate GKP state, the atomic state in the second encoding block is also optimized as given in Ref.~\cite{hastrupProtocolGeneratingOptical2022a}.

In Ref.~\cite{hastrupProtocolGeneratingOptical2022a}, the channel is represented by a Kraus map in the Fock basis. For three encoding blocks, a photon-number cutoff of 230 was used, and therefore the generated density matrix contains $D^2=231^2=53\,361$ entries, where $D$ is the Fock space dimension. By using the sparsity of the Kraus operator, the required computational cost is $\mathcal O(N_K D^2)$, where $N_K$ is the number of Kraus operators associated with the number of lost photons. Since the cavity photons are scattered to two different modes by intracavity loss and atomic scattering, $N_K$ is given by the combination of the lost photons to these two modes. If all combinations are taken into account,
\begin{equation}
    N_K=\sum_{n_\text{loss}=0}^{230}(n_\text{loss}+1)= 26\,796.
\end{equation}
However, losing a large number of photons rarely happens, so the required $N_K$ to approximate the output state is generally lower than $26\,796$. If one considers up to $N_\text{loss}$ photons, $N_K=(N_\text{loss}+1)(N_\text{loss}+2)/2$. $N_\text{loss}$ is numerically determined such that the probability of losing more than $N_\text{loss}$ photons is negligible. Generally, losses of tens of photons should be taken into account, leading to $10^2\lesssim N_K$. Therefore, the computational cost is $5\times 10^6\lesssim N_K D^2\lesssim 1.4\times 10^9$.

Conversely, in the hybrid LCoG representation, the memory cost only scales as $\mathcal O(N_LM^2)$ with the number of Gaussians $N_L$ and modes $M$. In this protocol, $N_L$ grows as $4^{N_\text{itr}}$ with the number of iterations $N_\text{itr}$. After the third encoding block, $N_L$ is only $64$, leading to $4N_LM^2=256$. More precisely, the total number of the entries in the log-weights, the mean vectors, and the covariance matrices is $64+2\times 64+2^2\times 64=448$. This is more than two orders of magnitude smaller than the entries of the Fock basis representation, $D^2=231^2=53\,361$. In addition, $\hat U_\text{cav}$ independently acts on each Gaussian term. Since it involves three modes, its computational cost scales as $\mathcal O(6^3 4^{N_\text{itr}})$. For $N_\text{itr}=3$, this estimate gives the computational cost $6^3 4^{N_\text{itr}}=6^3\times 64=13\,824$.  Thus, although the computational cost grows exponentially with $N_\text{itr}$, it is substantially smaller than that of the Fock basis representation.

\section{Conclusion}
In this paper, we proposed a novel simulation framework for CV--DV hybrid systems, including superconducting circuits, cavity-QED systems, trapped ions, hybrid optomechanics, and nonlinear optics. The framework represents a hybrid density matrix as a block matrix in the DV computational basis and describes each block by an LCoG expansion.
It supports DV, CV, and CV--DV hybrid operations, while the CV operations and CV--DV hybrid operations are restricted to Gaussian and controlled Gaussian operations, respectively. Under these assumptions, the CV part can be updated without a truncation of Fock basis, and the computational cost for updating each Gaussian term scales polynomially with the number of modes. This is a strong advantage over the previous simulators~\cite{stavengerC2QABosonicQiskit2022,mohapatraHyQBenchBenchmarkSuite2026a,furchesHybridlaneSoftwareDevelopment2026c}, whose computational costs scale exponentially with both the number of qubits and modes. 

By changing the computational basis into the Pauli basis, Clifford operations can be efficiently simulated. However, it imposes an additional computational cost to controlled Gaussian operations. Therefore, the choice of basis strongly depends on a simulated circuit. A Pauli-basis representation can be advantageous for circuits dominated by Clifford operations, while the computational basis is more suitable when controlled Gaussian operations are frequent.

The main technical results are the state representation and the transformation rule for controlled Gaussian gates. 
By dividing the hybrid state into the qubit-wise block matrices, the CV state in each block can be represented by LCoG formalism. 
For controlled Gaussian operations, different Gaussian transformations act on the CV part from the left and right, so the standard Gaussian-state update rule cannot be directly used. We showed that each complex Gaussian term nevertheless remains Gaussian and derived transformation rules for its log-weight, mean vector, and covariance matrix. The transformation is represented by joint matrices composed of symplectic matrices and the matrix representing left- and right-action.
We also extended these rules to controlled Gaussian channels with state-dependent Gaussian noise.

As an example, we simulated GKP state generation based on the cavity-QED system. The cavity interaction is treated as a controlled Gaussian channel acting on the input mode and two loss modes, and therefore this protocol is effectively described by the LCoG formalism. For three encoding blocks, our method requires 64 Gaussian terms, resulting in $448$ entries involving the log-weights, the mean vectors, and the covariance matrices, while the Fock-basis representation contains more than $5\times 10^4$ entries. Since the unitary independently acts on each Gaussian term, the computational cost is substantially lower than the Fock-basis expansion.

Our framework can be used for efficient simulation of versatile applications: non-Gaussian state generation, quantum sensing, quantum communication, quantum simulation, and hybrid quantum algorithms. Since it also supports noise channels, it can analyze the effect of decoherence in the above protocols for the various platforms, including superconducting circuits, cavity-QED systems, trapped ions, nonlinear optics, and hybrid optomechanical systems.

\section*{Data Availability}

The data that support the findings of this study are available from the corresponding author upon reasonable request.

\begin{acknowledgments}
This work was supported by JST Moonshot R\&D, Grant No. JPMJMS256E and JPMJMS256G, JST ASPIRE, Grant No. JPMJAP2427, JST COI-NEXT, Grant No. JPMJPF2221, Program for the Advancement of Next Generation Research Projects, and Keio University. We also acknowledge support from Danish National Research Foundation (bigQ, DNRF0142), EU ERC project ClusterQ (grant agreement no. 101055224, ERC-2021-ADG), EU (CLUSTEC, grant agreement no. 101080173, and QuantERA - ClusSTAR) and Innovation Fund Denmark (PhotoQ 3155-00024A and ScaleQ 5297-00043A). K.T. and S.K. were supported by JST SPRING, Grant No. JPMJSP2123.
\end{acknowledgments}

\appendix

\section{\label{appendix:Weyl_transform} Weyl transform of off-diagonal blocks}
In this appendix, we derive the transformation rule for off-diagonal blocks under controlled Gaussian unitaries. As discussed in the main text, we decompose the density matrix into blocks with respect to the qubit basis. A controlled Gaussian unitary acts as an ordinary Gaussian unitary on each diagonal block. On an off-diagonal block, however, different Gaussian unitaries act from the left and right, and the standard Gaussian transformation cannot be applied. Therefore, we derive the corresponding transformation rule below.

We consider an $M$-mode operator $\hat \varrho$, whose Weyl transform leads to Gaussian,
\begin{equation}
    W_{\hat \varrho}(\bm q^\prime)=e^{d}G_{\mu,\bm{\sigma}}(\bm q^\prime).
\end{equation} 
Suppose $\hat U_L$ and $\hat U_R$ denote the Gaussian unitaries acting on $\hat \varrho$ from left and right, respectively. The resulting state is  
\begin{equation}
    \hat \varrho_{LR}=\hat U_L \hat \varrho \hat U_R^\dagger.
\end{equation}
By definition in Eq.~\eqref{eq:Weyl_transform}, its Weyl transform is
\begin{align}
    \label{eq:Weyl_transform_of_off_diagonal_term}
    W_{\hat \varrho_{LR}}(\bm q)=\frac{1}{(2\pi^2)^M}\int d^{2M}\bm \zeta \Tr[\hat \Theta_{\bm q,\bm \zeta}^{LR}\hat \varrho]e^{i\bm q^T\Omega\bm \zeta},
\end{align}
where 
\begin{equation}
    \label{eq:Pi_LR}
    \hat \Theta_{\bm q,\bm \zeta}^{LR}=\hat U_R^\dagger \hat \Theta_{\bm q,\bm \zeta}\hat U_L.
\end{equation}
In the following subsections, we evaluate Eq.~\eqref{eq:Weyl_transform_of_off_diagonal_term} for controlled displacements, phase rotations, squeezing operations, and beam splitters. We label the target mode as mode $1$, except for a controlled beam splitter, whose target modes are labeled as $1$ and $2$.

\subsection{Controlled displacement gate}
We first consider a controlled displacement. The unitary operators in Eq.~(\ref{eq:Pi_LR}) are replaced with the Weyl operators as
\begin{equation}
    \begin{alignedat}{2}
        \hat U_L&= \hat D\left(\frac{\bm \alpha_L}{\sqrt 2}\right)
        &,\qquad
        \hat U_R&= \hat D\left(\frac{\bm \alpha_R}{\sqrt 2}\right),
    \end{alignedat}
\end{equation}
where $\bm \alpha_L$ and $\bm \alpha_R$ specify the left and right displacement vectors, respectively.
We obtain
\begin{align}
    \label{eq:Weyl_transform_CD_initial}
    W_{\hat \varrho_{LR}}(\bm q)=\frac{1}{(2\pi^2)^M}\int d^{2M}\bm \zeta \Tr[\hat \Theta_{\bm q,\bm \zeta}^{LR}\hat \varrho]e^{i\bm q^T\Omega\bm \zeta},
\end{align}
with
\begin{equation}
    \hat \Theta_{\bm q,\bm \zeta}^{LR}=\hat D^\dagger\left(\frac{\bm \alpha_R}{\sqrt 2}\right)\hat \Theta_{\bm q,\bm \zeta}\hat D\left(\frac{\bm \alpha_L}{\sqrt 2}\right).
\end{equation}
Using the relation,
\begin{equation}
    \hat D(\bm \beta)\hat D^\dagger(\bm \alpha)=\hat D(\bm \beta-\bm \alpha)\exp(i\bm \beta^T\Omega\bm \alpha),
\end{equation}
and the definition of $\hat \Theta_{\bm q,\bm \zeta}^{LR}$ in Eq.~(\ref{eq:definition_of_Pi_q_zeta}), we find
\begin{align}
    \hat \Theta_{\bm q,\bm \zeta}^{LR}=&\hat D\left(\frac{\bm q-\bm \zeta-\bm \alpha_R}{\sqrt 2}\right)\ket 0 \bra 0 \hat D^\dagger\left(\frac{\bm q+\bm \zeta-\bm \alpha_L}{\sqrt 2}\right)\nonumber\\
    \times& \exp[\frac{i}{2}\bm \alpha_-^T\bm \Omega\bm q-\frac{i}{2}\bm \alpha_{+}^T\bm \Omega\bm \zeta],
\end{align}
where $\bm \alpha_\pm=\bm\alpha_R\pm\bm \alpha_L$. Using Eq.~(\ref{eq:params_outer_product_of_coherent_state}), the Weyl transform of $\hat \Theta_{\bm q,\bm \zeta}^{LR}$ is given by
\begin{align}
    W_{\hat \Theta_{\bm q,\bm \zeta}^{LR}}(\bm q^\prime)=e^{\delta}G_{\bm \nu, \bm \omega}(\bm q^\prime),
\end{align}
where
\begin{subequations}
    \begin{align}
    \bm \nu=&\Pi_0(\mathbb I)(\bm q-\bm \alpha_R)+\Pi_1(\mathbb I)(\bm q-\bm \alpha_L)-i\Omega \bm \zeta\\
    \bm \omega&=\frac{1}{2}\mathbb I\\
    \delta=&(\bm q-\bm \alpha_R-\bm\zeta)^T\Pi_1(\mathbb I)(\bm q-\bm \alpha_L+\bm \zeta)\nonumber\\
    -&\frac{\|\bm q-\bm \alpha_R-\bm \zeta\|^2+\|\bm q-\bm \alpha_L+\bm \zeta\|^2}{4}\nonumber\\
    +&\frac{i}{2}\bm \alpha_-^T\bm \Omega\bm q-\frac{i}{2}\bm \alpha_+^T\bm \Omega\bm \zeta,
\end{align}
\end{subequations}
with $\Pi_s(\mathbb I)=\frac{1}{2}\left[\mathbb I+(-1)^s i\bm \Omega\right]$.
Since the Weyl transforms of both $\hat \Theta_{\bm q,\bm \zeta}^{LR}$ and $\hat \varrho$ are Gaussian, the trace reads
\begin{align}
    \Tr[\hat \Theta_{\bm q,\bm \zeta}^{LR}\hat \varrho]=(2\pi)^Me^{d+\delta}G_{\bm \mu,\bm \Sigma}(\bm \nu),
\end{align}
where $\bm \Sigma=\bm{\sigma}+\bm \omega$.
Inserting this equation into Eq.~(\ref{eq:Weyl_transform_CD_initial}), we obtain
\begin{align}
    W_{\hat \varrho_{LR}}(\bm q)=\frac{1}{\pi^M}\int d^{2M}\bm \zeta\ G_{\bm \mu,\bm \Sigma}(\bm \nu)e^{i\bm q^T\bm \Omega\bm \zeta+\delta+d},
\end{align}
The exponent in the integrand can be transformed into a quadratic form with respect to $\bm \zeta$:
\begin{align}
    \label{eq:Quadratic_form}
    W_{\hat \varrho_{LR}}(\bm q)=\frac{1}{\sqrt{\det 2\pi^2 \Sigma}}\int d^{2M}\bm \zeta\ e^{-\frac{1}{2}\bm \zeta^T A \bm \zeta +\bm b^T \bm \zeta + c},
\end{align}
where
\begin{equation}
    \begin{aligned}
        A=&\bm \Omega^T(2\mathbb I-C_2)\bm \Omega,\\
        \bm b^T=&\bm q^T B_1^T+\bm B_0^T,\\
        c=&-\frac{1}{2}\bm q^T C_2 \bm q+\bm C_1^T\bm q+\frac{1}{2}C_0,
    \end{aligned}
\end{equation}
with
\begin{equation}
    \begin{aligned}
        \bm B_0^T=&-i\bm C_1^T\bm \Omega,\\
        B_1^T=&iC_2\bm \Omega,\\
        C_0=&-\bm \eta^T\bm \Sigma^{-1}\bm \eta+2d,\\
        &-\bm\alpha_-^T[-\Pi_1(\mathbb I)\bm \alpha_L+\Pi_0(\mathbb I)\bm \alpha_R],\\
        \bm C_1^T=&\bm \eta^T\bm \Sigma^{-1}+i\bm \alpha_-^T\Omega,\\
        C_2=&\bm \Sigma^{-1},\\
        \bm \eta=&\Pi_1(\mathbb I)\bm \alpha_L+\Pi_0(\mathbb I)\bm \alpha_R+\bm \mu.
    \end{aligned}
\end{equation}
Evaluating the Gaussian integral yields
\begin{align}
    W_{\hat \varrho_{LR}}(\bm q)=\frac{1}{\pi^M\sqrt{\det \Sigma A}}\exp\left(\frac{1}{2}\bm b^TA^{-1}\bm b+c\right).
\end{align}
Completing the square in $\bm q$, we write the exponent:
\begin{equation}
    \begin{aligned}
        &\frac{1}{2}\bm b^TA^{-1}\bm b+c\\
        =&-\frac{1}{2}\left(\bm q-\bm \mu_\text{out}\right)^T\bm{\sigma}_\text{out}^{-1}\left(\bm q-\bm \mu_\text{out}\right)+\bm\Delta,
    \end{aligned}
\end{equation}
Matching the quadratic, linear, and constant terms on both sides gives
\begin{equation}
    \label{eq:coefficients_in_output_and_ABC}
    \begin{aligned}
        -\bm{\sigma}_\text{out}^{-1}=&B_1^TA^{-1}B_1-C_2\\
        \bm\mu_\text{out}^T\bm{\sigma}_\text{out}^{-1}=&\bm B_0^TA^{-1}B_1+\bm C_1^T\\
        \bm\Delta-\frac{1}{2}\bm\mu_\text{out}^T\bm{\sigma}_\text{out}^{-1}\bm \mu_\text{out}=&\frac{1}{2}\bm B_0^T A^{-1}\bm B_0+\frac{1}{2}C_0.
    \end{aligned}
\end{equation}
Solving these relations results in
\begin{equation}
    \label{eq:coefficients_in_output_and_C}
    \begin{aligned}
        \bm{\sigma}_\text{out}&=C_2^{-1}-\frac{1}{2}\mathbb I\\
        \bm \mu_\text{out}&=C_2^{-1}\bm C_1\\
        \bm \Delta&=\frac{1}{2}\bm C_1^TC_2^{-1}\bm C_1+\frac{1}{2}C_0.
    \end{aligned}
\end{equation}
From the normalization condition of the Gaussian function, the log-weight is defined by
\begin{equation}
    e^{d_\text{out}}=e^\Delta\sqrt{\frac{\det 2\bm{\sigma}_\text{out}}{\det (\Sigma A)}}.
\end{equation}
After some algebra, we find
\begin{subequations}
    \begin{align}
        \bm \mu_\text{out}=&\bm \mu+\mathcal{\bm A}_{LR}\\
        \bm{\sigma}_\text{out}=&\bm{\sigma}\\
        d_\text{out}=&d-\frac{i}{2}(\bm \alpha_L-\bm \alpha_R)^T\Omega\left(\bm{\mathcal{A}}_{LR}+2\bm\mu\right),
    \end{align}
\end{subequations}
where $\mathcal{\bm A}_{LR}$ denotes a joint-displacement vector:
\begin{equation}
    \mathcal{\bm A}_{LR}=\Pi_0(2\bm{\sigma})\bm \alpha_L+\Pi_1(2\bm{\sigma})\bm \alpha_R,
\end{equation}

\subsection{\label{appendix:controlled_phase_rotation_gate}Controlled phase-rotation gate}
We next consider a controlled phase-rotation gate. 
The unitary operators in Eq.~(\ref{eq:Pi_LR}) are replaced with phase-rotation operators as
\begin{equation}
    \begin{alignedat}{2}
        \hat U_L&=\hat R\left(\theta_L \right),
        &\qquad
        \hat U_R&= \hat R\left(\theta_R\right).
    \end{alignedat}
\end{equation}
Inserting them into Eq.~(\ref{eq:Pi_LR}) leads to
\begin{align}
    \label{eq:Pi_LR_phase}
    \hat \Theta_{\bm q,\bm \zeta}^{LR}&=\hat R^\dagger(\theta_R)\hat \Theta_{\bm q,\bm \zeta}\hat R(\theta_L)\nonumber\\
    &=\hat D\left(\mathsf{R}_R^{-1}\frac{\bm q-\bm \zeta}{\sqrt 2}\right)\ket{0}\bra{0}\hat D^\dagger\left(\mathsf{R}_L^{-1}\frac{\bm q+\bm \zeta}{\sqrt 2}\right),
\end{align}
where $\mathsf{R}_{L}$ and $\mathsf{R}_{R}$ denote the symplectic matrices corresponding to the phase-rotation operator $\hat R(\theta_{L})$ and $\hat R(\theta_{R})$, respectively. Since $\hat \Theta_{\bm q,\bm \zeta}^{LR}$ is the outer product of the two different coherent states, the Weyl transform of $\hat \Theta_{\bm q,\bm \zeta}^{LR}$ is given by
\begin{align}
    W_{\hat \Theta_{\bm q,\bm \zeta}^{LR}}(\bm q^\prime)=e^{\delta}G_{\bm \nu, \bm \omega}(\bm q^\prime),
\end{align}
where
\begin{subequations}
    \begin{align}
        \bm \nu=&\Pi_0(\mathbb I)\mathsf{R}_{R}^{-1}\bm q+\Pi_1(\mathbb I)\mathsf{R}_{L}^{-1}\bm q-i\bm \Omega P_{\mathsf{R}}\bm \zeta,\\
        \delta=&(\bm q-\bm\zeta)^T\mathsf{R}_{R}\Pi_1(\mathbb I)\mathsf{R}_{L}^{-1}(\bm q+\bm \zeta)\nonumber\\
        &-\frac{\|\bm q\|^2+\|\bm\zeta\|^2}{2},
    \end{align}
\end{subequations}
where $P_{\mathsf{R}}=\Pi_0(\mathbb I)\mathsf{R}_{R}^{-1}+\Pi_1(\mathbb I)\mathsf{R}_{L}^{-1}$.
As in the previous section, the trace reduces to a Gaussian function in phase space as 
\begin{equation}
    \Tr[\hat \Theta_{\bm q,\bm \zeta}^{LR}\hat \varrho]=(2\pi)^Me^{d+\delta}G_{\bm \mu,\bm \Sigma}(\bm \nu).
\end{equation}
Then, the Weyl transform of $\hat \varrho_{LR}$ is given by the following Gaussian integral:
\begin{align}
    W_{\hat \varrho_{LR}}(\bm q)=\frac{1}{\sqrt{\det 2\pi^2 \Sigma}}\int d^{2M}\bm \zeta\ e^{-\frac{1}{2}\bm \zeta^T A \bm \zeta +\bm b^T \bm \zeta + c},
\end{align}
where
\begin{equation}
    \begin{aligned}
        A=&\bm \Omega^T(2\mathbb I-C_2)\bm \Omega,\\
        \bm b^T=&\bm q^T B_1^T+\bm B_0^T,\\
        c=&-\frac{1}{2}\bm q^T C_2 \bm q+\bm C_1^T\bm q+\frac{1}{2}C_0,
    \end{aligned}
\end{equation}
with
\begin{equation}
    \begin{aligned}
        \bm B_0^T=&-i\bm C_1^T\bm \Omega,\\
        B_1^T=&iC_2\bm \Omega,\\
        C_0=&-\bm \mu^T\bm \Sigma^{-1}\bm \mu+2d,\\
        \bm C_1^T=&\bm \mu^T\bm \Sigma^{-1}P_{\mathsf{R}},\\
        C_2=&P_{\mathsf{R}}^T\bm \Sigma^{-1}P_{\mathsf{R}}+\mathbb I-P_{\mathsf{R}}^T P_{\mathsf{R}}.
    \end{aligned}
\end{equation}
The Gaussian integral reads
\begin{align}
    W_{\hat \varrho_{LR}}(\bm q)=\frac{1}{\pi^M\sqrt{\det \Sigma A}}\exp\left(\frac{1}{2}\bm b^TA^{-1}\bm b+c\right).
\end{align}
For the controlled phase-rotation gate, the relation in Eq.~(\ref{eq:coefficients_in_output_and_C}) holds and results in
\begin{subequations}
    \begin{align}
        \bm \mu_\text{out}=&\mathcal{R}_{LR}^{-1}(2\bm{\sigma})\bm \mu,\\
        \bm{\sigma}_\text{out}=&\mathcal{R}_{LR}^{-1}(2\bm{\sigma})\bm{\sigma}\mathcal{R}_{LR}(\bm{\sigma}^{-1}/2),\\
        d_\text{out}=& d-\frac{i}{2}\bm \mu^T\left(\mathsf{R}_L^T-\mathsf{R}_R^T\right)\Omega\bm \mu_\text{out}+\mathcal N\\
        \mathcal N=&\ln \sqrt \frac{\det \bm{\sigma}_\text{out}+\mathbb I/2}{\det \bm{\sigma}+\mathbb I/2},
    \end{align}
\end{subequations}
where $\mathcal{R}_{LR}(\bm{\sigma})$ denotes the joint phase-rotation matrix:
\begin{align}
    \mathcal R_{LR}(\bm{\sigma})=\Pi_0(\bm{\sigma})\mathsf{R}_L^T+\Pi_1(\bm{\sigma})\mathsf{R}_R^T.
\end{align}

\subsection{\label{appendix:controlled_squeezing} Controlled squeezing gate}
We consider a controlled squeezing gate. The unitary operators in Eq.~(\ref{eq:Pi_LR}) are replaced with the squeezing operators as
\begin{equation}
    \begin{alignedat}{2}
        \hat U_L&= \hat S\left(r_L \right),
        &\qquad
        \hat U_R&=\hat S\left(r_R\right).
    \end{alignedat}
\end{equation}
Inserting them into Eq.~(\ref{eq:Pi_LR}) leads to
\begin{align}
    &\hat \Theta_{\bm q,\bm \zeta}^{LR}=\hat S^\dagger(r_R)\hat \Theta_{\bm q,\bm \zeta}\hat S(r_L)\nonumber\\
    =&\hat D\left(\mathsf{S}_R^{-1}\frac{\bm q-\bm \zeta}{\sqrt 2}\right)\ket{r_R}\bra{r_L}\hat D^\dagger\left(\mathsf{S}_L^{-1}\frac{\bm q+\bm \zeta}{\sqrt 2}\right),
\end{align}
where $\ket{r_R}\bra{r_L}=\hat S^\dagger(r_R)\ket{0}\bra{0}\hat S(r_L)$, and $\mathsf{S}_{L}$ and $\mathsf S_{R}$ denote the symplectic matrices corresponding to the squeezing operator $\hat S(r_{L})$ and $\hat S(r_{R})$, respectively. 

To derive the Weyl transform of $\hat \Theta_{\bm q,\bm \zeta}^{LR}$, we consider the Weyl transform of the outer product of the two different squeezed states $\ket{r_R}\bra{r_L}$:
\begin{align}
    W_{\ket{r_R}\bra{r_L}}(\bm q)=\frac{1}{(2\pi^2)^M}\int d^{2M}\bm \zeta \bra{r_L}\hat \Theta_{\bm q,\bm \zeta}\ket{r_R}e^{i\bm q^T\Omega\bm \zeta}.
\end{align}
Here, we assume that the first mode is squeezed, and the other $M-1$ modes remain in vacuum. Thus, we separately consider the first and the rest of modes. The Weyl transform at the first mode is
\begin{equation}
    \label{eq:Weyl_transform_of_rR_rL}
    W_{\ket{r_R}_1\bra{r_L}}(\bm q_1)=\frac{1}{2\pi^2}\int d^{2}\bm \zeta_1\ {}_{1}\bra{r_L}\hat \Theta_{\bm q_1,\bm \zeta_1}\ket{r_R}_1e^{i\bm q_1^T\Omega\bm \zeta_1}.
\end{equation}
The overlap between a coherent state and a squeezed state is given by 
\begin{align}
    \label{eq:overlap_D_and_S}
    {}_1\bra 0 \hat D\left(\frac{\bm q_1}{\sqrt 2}\right)\hat S(r)\ket{0}_1=\frac{1}{\sqrt{\cosh r}}\exp\left[-\frac{1}{2}\bm q_1^T \Xi \bm q_1\right],
\end{align}
where
\begin{align}
    \Xi=\frac{1}{2}\left(\begin{matrix}
        1+\tanh r&-i\tanh r\\
        -i\tanh r & 1-\tanh r
    \end{matrix}\right).
\end{align}
Since Eq.~(\ref{eq:overlap_D_and_S}) is Gaussian function, Eq.~(\ref{eq:Weyl_transform_of_rR_rL}) can be analytically evaluated through Gaussian integral:
\begin{equation}
    W_{\ket{r_R}_1\bra{r_L}}(\bm q_1)=\frac{1}{\sqrt{\cosh{(r_R-r_L)}}}G_{0,\bm \varsigma}(\bm q_1),
\end{equation}
where the covariance matrix $\bm \varsigma$ is given by
\begin{align}
    \bm \varsigma=\frac{(1+\lambda_L\lambda_R)\mathbb I+2Z\left(\lambda_R\Pi_1+\lambda_L\Pi_0\right)}{2(1-\lambda_L\lambda_R)},
\end{align}
with $\lambda_{L}=\tanh(r_{L})$, $\lambda_{R}=\tanh(r_{R})$, and $Z=\text{diag}(1,-1)$.
Then, the Weyl transform of $\hat \Theta_{\bm q,\bm \zeta}^{LR}$ is then given by
\begin{align}
    W_{\hat \Theta_{\bm q,\bm \zeta}^{LR}}(\bm q^\prime)=e^{\delta}G_{\bm \nu,\bm \varsigma\oplus \bm \omega_{M-1}}(\bm q^\prime),
\end{align}
where $\bm \omega_{M-1}$ represents the covariance matrix of the vacuum state in $M-1$ modes, and the complex mean vector and the log-weight are given by
\begin{equation}
    \begin{aligned}
        \bm \nu&=\Pi_0(2\varsigma)\mathsf{S}_{R}^{-1}(\bm q-\bm \zeta)+\Pi_1(2\varsigma)\mathsf{S}_{L}^{-1}(\bm q+\bm \zeta),\\
        e^{\delta}&=\frac{1}{\sqrt{\cosh{(r_L-r_R)}}}e^{-\frac{i}{2}[\mathsf{S}_{R}^{-1}(\bm q-\bm \zeta)-\mathsf{S}_{L}^{-1}(\bm q+\bm \zeta)]^T\bm \Omega\bm\nu}.
    \end{aligned}
\end{equation}

Since $W_{\hat{\Theta}_{\bm q,\bm \zeta}^{LR}}(\bm q^\prime)$ is Gaussian, the Weyl transform of $\hat \varrho_{LR}$ is derived through the same procedure in Appendix~\ref{appendix:controlled_phase_rotation_gate}. Trace in Eq.~(\ref{eq:Weyl_transform_of_off_diagonal_term}) is
\begin{equation}
    \Tr[\hat \Theta_{\bm q,\bm \zeta}^{LR}\hat \varrho]=(2\pi)^M e^{d+\delta}G_{\bm\mu,\bm\Sigma^\prime}(\bm\nu),
\end{equation}
where $\bm\Sigma^\prime=\bm{\sigma}+(\bm\varsigma\oplus \bm \omega_{M-1})$. Then, the Weyl transform of $\hat \varrho_{LR}$ is given by the following quadratic form:
\begin{align}
    W_{\hat \varrho_{LR}}(\bm q)=\frac{1}{\sqrt{\det 2\pi^2 \Sigma}}\int d^{2M}\bm \zeta\ e^{-\frac{1}{2}\bm \zeta^T A \bm \zeta +\bm b^T \bm \zeta + c},
\end{align}
with
\begin{equation}
    \begin{aligned}
        A=&\bm \Omega^T(2\mathbb I-C_2)\bm \Omega,\\
        \bm b^T=&\bm q^T B_1^T+\bm B_0^T,\\
        c=&-\frac{1}{2}\bm q^T C_2 \bm q+\bm C_1^T\bm q+\frac{1}{2}C_0,
    \end{aligned}
\end{equation}
where 
\begin{equation}
    \begin{aligned}
        \bm B_0^T=&-i\bm C_1^T\bm \Omega,\\
        B_1^T=&iC_2\bm \Omega,\\
        C_0=&-\bm \mu^T{\bm \Sigma^\prime}^{-1}\bm \mu+2d-\ln \cosh{(r_L-r_R)},\\
        \bm C_1^T=&\bm \mu^T{\bm \Sigma^\prime}^{-1}P_{\mathsf{S}},\\
        C_2=&P_{\mathsf{S}}^T{\bm \Sigma^\prime}^{-1}P_{\mathsf{S}}+i\text{Sym}\left((\mathsf S_R^{-1}-\mathsf S_L^{-1})^T\bm \Omega P_{\mathsf{S}}\right)
    \end{aligned}
\end{equation}
Here, $P_{\mathsf{S}} = \Pi_0(2\bm \varsigma)\mathsf S_R^{-1}+\Pi_1(2\bm \varsigma)\mathsf S_L^{-1}$, and $\text{Sym}(A)=\frac{1}{2}(A+A^T)$ indicates the symmetrization of a matrix $A$.
The Gaussian integral results in
\begin{align}
    W_{\hat \varrho_{LR}}(\bm q)=\frac{1}{\pi^M\sqrt{\det \Sigma^\prime A}}\exp\left(\frac{1}{2}\bm b^TA^{-1}\bm b+c\right).
\end{align}
As in the controlled phase-rotation gate, the relation in Eq.~(\ref{eq:coefficients_in_output_and_C}) holds. Therefore, we find
\begin{subequations}
    \begin{align}
        \bm \mu_\text{out}=&\mathcal{S}_{LR}^{-1}(2\bm{\sigma})\bm \mu,\\
        \bm{\sigma}_\text{out}=&\mathcal{S}_{LR}^{-1}(2\bm{\sigma})\bm{\sigma}\mathcal{S}_{LR}^{[T]}(\bm{\sigma}^{-1}/2),\\
        d_\text{out}=& d-\frac{i}{2}\bm \mu^T\left(\mathsf{S}_L^T-\mathsf{S}_R^T\right)\bm\Omega\bm \mu_\text{out}+\mathcal{N}_\text{sq}\\
        \mathcal N_\text{sq}=&\ln \sqrt \frac{\det (\bm{\sigma}_\text{out}+\mathbb I/2)}{\cosh(r_L-r_R)\det \Sigma^\prime},
    \end{align}
\end{subequations}
where $\mathcal S_{LR}(\bm{\sigma})$ and $\mathcal S_{LR}^{[T]}(\bm{\sigma})$ are the joint squeezing matrices,
\begin{equation}
    \label{eq:joint_squeezing_LR}
    \begin{aligned}
        \mathcal S_{LR}(\bm{\sigma})&=\Pi_0(\bm{\sigma})\mathsf{S}_L^{-1}+\Pi_1(\bm{\sigma})\mathsf{S}_R^{-1},\\
        \mathcal S_{LR}^{[T]}(\bm{\sigma})&=\Pi_0(\bm{\sigma})\mathsf{S}_L^{T}+\Pi_1(\bm{\sigma})\mathsf{S}_R^{T}.
    \end{aligned}
\end{equation}

\subsection{Controlled beam splitter gate}
In this section, we consider the controlled beam splitter gate. 
The unitary operators in Eq.~(\ref{eq:Pi_LR}) are replaced with the beam splitter operators as
\begin{equation}
    \hat U_L\rightarrow \hat B\left(\varphi_L \right),\qquad\hat U_R\rightarrow \hat B\left(\varphi_R\right).
\end{equation}
Inserting them into Eq.~(\ref{eq:Pi_LR}) leads to
\begin{align}
    \label{eq:Theta_LR_BS}
    \hat \Theta_{\bm q,\bm \zeta}^{LR}&=\hat B^\dagger(\varphi_R)\hat \Theta_{\bm q,\bm \zeta}\hat B(\varphi_L)\nonumber\\
    &=\hat D\left(\mathsf{B}_R^{-1}\frac{\bm q-\bm \zeta}{\sqrt 2}\right)\ket{0}\bra{0}\hat D^\dagger\left(\mathsf{B}_L^{-1}\frac{\bm q+\bm \zeta}{\sqrt 2}\right),
\end{align}
where $\mathsf{B}_{L}$ and $\mathsf{B}_{R}$ denote the symplectic matrix corresponding to the beam splitter operator $\hat B(\varphi_{L})$ and $\hat B(\varphi_{R})$, respectively. Since beam splitter operator is a passive unitary operator as well as a phase rotation operator,  Eq.~\eqref{eq:Theta_LR_BS} has the same form as Eq.~(\ref{eq:Pi_LR_phase}). Following the derivation in Appendix~\ref{appendix:controlled_phase_rotation_gate}, we find
\begin{equation}
    \begin{aligned}
        \bm \mu_\text{out}=&\mathcal{B}_{LR}^{-1}(2\bm{\sigma})\bm \mu,\\
        \bm{\sigma}_\text{out}=&\mathcal{B}_{LR}^{-1}(2\bm{\sigma})\bm{\sigma}\mathcal{B}_{LR}(\bm{\sigma}^{-1}/2),\\
        d_\text{out}=& d-\frac{i}{2}\bm \mu^T\left(\mathsf{B}_L^T-\mathsf{B}_R^T\right)\Omega\bm \mu_\text{out}\\
        +&\ln \sqrt \frac{\det \bm{\sigma}_\text{out}+\mathbb I/2}{\det \bm{\sigma}+\mathbb I/2},
    \end{aligned}
\end{equation}
where $\mathcal{B}_{LR}(\bm{\sigma})$ denotes the joint beam splitter matrix, 
\begin{align}
    \mathcal B_{LR}(\bm{\sigma})=\Pi_0(\bm{\sigma})\mathsf{B}_L^T+\Pi_1(\bm{\sigma})\mathsf{B}_R^T.
\end{align}

\section{\label{appendix:cQED_controlled_unitary}Derivation of controlled unitaries in cavity-QED systems}
In this section, we derive symplectic matrices corresponding to the cavity-QED system in Fig.~\ref{fig:cavity-QED_setup}(a).
We first consider the dynamics of intracavity photons. The cavity mode $\hat a_c$ couples to the external output field $\hat a_\text{out}$ at rate $\kappa_\text{ex}$ and leaks to the mode $\hat b_\text{out}$ at rate $\kappa_\text{in}$, while the system evolves under the Hamiltonian $\hat H$. The Heisenberg--Langevin equation is given by
\begin{equation}
    \label{eq:HL_equation_cavity}
    \begin{aligned}
        \dot{\hat a}_c=-i[\hat a_c,\hat H]-\sqrt{2\kappa_\text{ex}}\hat a_\text{out}-\sqrt{2\kappa_\text{in}}\hat b_\text{out}+\kappa \hat a_c,
    \end{aligned}
\end{equation}
where $\kappa=\kappa_\text{in}+\kappa_\text{ex}$ denotes the total decay rate of the cavity. Since the atom-cavity coupling depends on the atomic states, we first consider the case of the atom in $\ket{1}$. The system Hamiltonian is then given by 
\begin{equation}
    \hat H=\omega_a \hat \sigma^\dagger \hat \sigma +\omega_c \hat a_c^\dagger \hat a_c+g(\hat a_c^\dagger \hat \sigma+\hat \sigma^\dagger\hat a_c),
\end{equation}
where $\omega_a$ represents the transition frequency between $\ket e$ and $\ket 1$, $\omega_c$ denotes the cavity resonance frequency, $g$ denotes the atom--cavity coupling, and $\hat{\sigma}=\ket{1}\bra{e}$ is the ladder operator. To solve Eq.~(\ref{eq:HL_equation_cavity}), we also consider the equation of motion for $\hat \sigma$. As the excited state decays to the mode $\hat c_\text{out}$ at rate $\gamma$, its Heisenberg--Langevin equation is given by 
\begin{align}
    \label{eq:HL_equation_ladder}
    \dot{\hat \sigma}=-i[\hat \sigma,\hat H]+\hat \sigma_z\left(-\gamma \hat \sigma+\sqrt{2\gamma}\hat c_\text{out}\right),
\end{align}
where $\hat \sigma_z=-\ket{1}\bra{1}+\ket{e}\bra{e}$. 
Under the weak excitation limit $\langle \hat \sigma_z \rangle\sim -1$, we replace $\hat \sigma_z$ with $-1$ and obtain the linearized Heisenberg-Langevin equation,
\begin{align}
    \label{eq:HL_equation_ladder_approx}
    \dot{\hat \sigma}=-i[\hat \sigma,\hat H]+\gamma \hat \sigma-\sqrt{2\gamma}\hat c_\text{out}.
\end{align}

To obtain the three-mode unitary shown in Fig.~\ref{fig:cavity_QED_channel}, we use the input-output relation: 
\begin{equation}
    \begin{aligned}
        \hat a_\text{out}&=\hat a_\text{in}+\sqrt{2\kappa_\textbf{ex}}a_c,
    \end{aligned}
\end{equation}
where $\hat a_\text{in}$ represents the input fields of the cavity. In addition, we also consider the input-output relations corresponding to the intracavity photon loss and photon scattering as
\begin{equation}
    \begin{aligned}
        \hat b_\text{out}&=\hat b_\text{in}+\sqrt{2\kappa_\textbf{in}}a_c,\\
        \hat c_\text{out}&=\hat c_\text{in}+\sqrt{2\gamma}\hat \sigma,\\
    \end{aligned}
\end{equation}
where $\hat b_\text{in}$ and $\hat c_\text{in}$ denote the input fields corresponding to loss and spontaneous decay. 

By moving to the Fourier domain, the above equations reduce to the following input-output relation at resonant condition $\omega_a=\omega_c$: 
\begin{widetext}
    \begin{align}
        \begin{bmatrix}
            \hat a_\text{out}\\
            \hat b_\text{out}\\
            \hat c_\text{out}
        \end{bmatrix}=\frac{1}{1+2C}\begin{bmatrix}
            1+2C-2\eta&-2\sqrt{\eta(1-\eta)}&2i\sqrt{2\eta C}\\
            -2\sqrt{\eta(1-\eta)}&-1+2C+2\eta&2i\sqrt{2(1-\eta) C}\\
            2i\sqrt{2\eta C}&2i\sqrt{2(1-\eta) C}&-1+2C
        \end{bmatrix}\begin{bmatrix}
            \hat a_\text{in}\\
            \hat b_\text{in}\\
            \hat c_\text{in}
        \end{bmatrix},
    \end{align}
\end{widetext}
where $C=g^2/(2\kappa\gamma)$ denotes the cooperativity, and $\eta=\kappa_\text{ex}/\kappa$ denotes the escape efficiency. 
When the atom is in $\ket 0$, the atom-cavity coupling becomes $g=0$, leading to
\begin{align}
        \begin{bmatrix}
            \hat a_\text{out}\\
            \hat b_\text{out}\\
            \hat c_\text{out}
        \end{bmatrix}=\begin{bmatrix}
            1-2\eta&-2\sqrt{\eta(1-\eta)}&0\\
            -2\sqrt{\eta(1-\eta)}&-1+2\eta&0\\
            0&0&-1
        \end{bmatrix}\begin{bmatrix}
            \hat a_\text{in}\\
            \hat b_\text{in}\\
            \hat c_\text{in}
        \end{bmatrix}.
\end{align}
From the above equations, the symplectic matrices $\mathsf{U}(0)$ and $\mathsf{U}(g)$ corresponding to the unitary operators $\hat U(0)$ and $\hat U(g)$ in Eq.~(\ref{eq:controlled_phase_based_on_cavity_QED}) are given by
\begin{widetext}
\begin{equation}
    \label{eq:symplectic_matrices_in_cavity_QED_appendix}
    \begin{aligned}
        \mathsf{U}(0)&=
        \begin{bmatrix}
            (1-2\eta)I&-2\sqrt{\eta(1-\eta)}I&0\\
            -2\sqrt{\eta(1-\eta)}I&(-1+2\eta)I&0\\
            0&0&-I
        \end{bmatrix},\\
        \mathsf{U}(g)&=\frac{1}{1+2C}
        \begin{bmatrix}
            (1+2C-2\eta)I&-2\sqrt{\eta(1-\eta)}I&-2\sqrt{2\eta C}\,iY\\
            -2\sqrt{\eta(1-\eta)}I&(-1+2C+2\eta)I &-2\sqrt{2(1-\eta)C}iY\\
            -2\sqrt{2\eta C}iY&-2\sqrt{2(1-\eta)C}iY&(-1+2C)I
        \end{bmatrix},
    \end{aligned}
\end{equation}
\end{widetext}
where 
\begin{subequations}
    \begin{align}
        I&=\begin{bmatrix}
            1 & 0\\
            0 & 1
        \end{bmatrix},
        &\qquad
        Y&=\begin{bmatrix}
            0 & -i\\
            i & 0 
        \end{bmatrix}.
    \end{align}
\end{subequations}

\bibliography{Hybrid_LCoG}
\end{document}